\documentstyle[preprint,aps,eqsecnum,tighten,epsf,floats]{revtex}
\begin{document}

\draft
\preprint{
 \parbox{1.5in}{\leftline{JLAB-THY-98-32}
                \leftline{WM-98-109} }  }
\title{The Stability of the Spectator, Dirac, and Salpeter Equations for Mesons}

\author{Michael Uzzo}
\address{College of William and Mary, Williamsburg, Virginia 23187}

\author{Franz Gross}
\address{College of William and Mary, Williamsburg, Virginia 23187 \protect\\
and Thomas Jefferson National Accelerator Facility, \protect\\ 12000 Jefferson
Avenue, Newport News, Virginia 23606}

\date{\today}

\maketitle

\begin{abstract}
Mesons are made of quark-antiquark pairs held together by the 
strong force.  The one channel spectator, Dirac, and Salpeter equations can each
be used to model this pairing.  We look at cases where the relativistic kernel
of  these equations corresponds to a time-like vector exchange, a scalar
exchange,  or a linear combination of the two.  Since the model used in this
paper describes mesons which cannot decay physically, the equations must
describe stable states. We find that this requirement is not always
satisfied, and give a complete discussion of the conditions under which the
various equations give unphysical, unstable solutions.
\end{abstract}

\widetext

\section{Introduction}
\subsection{Background}

In the simplest models, mesons are bound states of a valance quark-antiquark pair
confined by the strong force.  Even for such a simple case a covariant model
is needed when the mesons are composed of light quarks with high momentum
components.  However, covariant models require knowledge of the
Lorentz structure of the confining interaction, and it turns out that
some choices of Lorentz structure for some equations will produce mesons 
which decay.  When no mechanism for decay has been included in the model (which
will be the situation for the cases discussed in this paper), this is a sign
that the solutions are unphysical.  In may be acceptable for an equation to
produce unstable (i.e. unphysical) solutions {/it if\/} these solutions are confined 
to a region of the spectrum which can be precisely characterized and systematically
ignored, but if this is not possible equations which produce such
unphysical solutions are unsatisfactory.  In this paper we  study
confining potentials with scalar and time-like vector exchanges, and find that
the stability of such interactions depends on the kind of relativistic
equation used for the description of the interaction.   
 
This is not the first time that the stability of covariant models of confinement
has been addressed. Several papers have been written on this topic,
some with contradictory conclusions. Two examples which illustrate this
are papers titled {\it An exact argument against an effective vector exchange
for the confining quark-antiquark potential\/} \cite{dg}, and {\it Evidence
against a scalar  confining potential in QCD\/} \cite{jfl}.  If both papers
are correct, this would indicate that, at best, the Lorentz structure for the
potential is more complex than a simple scalar or vector exchange.

Our research into the question of stability was motivated by the paper
of Parramore and Piekarewicz\cite{pp}, which found that the Salpeter equation
was stable when the vector strength exceeded the scalar strength.  This seemed
counter intuitive to us, since it is well known that, because of the famous
Klein paradox, the Dirac equation is stable only when the potential is
predominately scalar.  Their result also contradicted the work of
another group \cite{munz} who found that the Salpeter equation was stable for
a pure scalar confining interaction, provided the quark mass was sufficiently
large.

%
%
\begin{figure}
\begin{center}
\mbox{
   \epsfysize=1.5in
\epsfbox{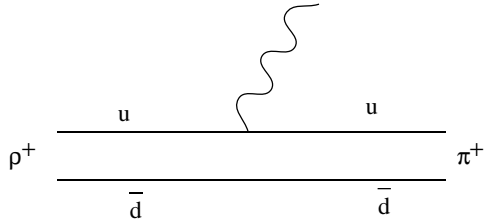}
}
\end{center}
\caption{Example of electroweak decay of the $\rho^{+}$ meson.}
\label{figa}
\end{figure}

%
%
\begin{figure}
\begin{center}
\mbox{
   \epsfysize=1.5in
\epsfbox{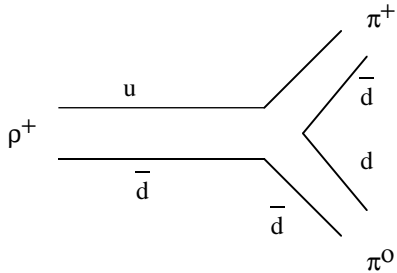}
}
\end{center}
\caption{Example of strong decay of the $\rho^{+}$ meson.}
\label{figb}
\end{figure}

\subsection{Physical and unphysical instabilities}   

We begin the discussion by making a distinction between instabilities which
are physical and those which are unphysical. Real mesons have a finite
lifetime and can decay either through the strong interaction or the
electroweak interaction.  For example, the $\rho^{+}$ can decay into a photon
and a $\pi^{+}$ through the electroweak interaction shown in Fig.~\ref{figa}. 
It can also decay into  a $\pi^{+}$ and $\pi^{0}$ via the strong interaction,
as shown in Fig.~\ref{figb}. In this paper we describe mesons which are
isolated from external influences (including vacuum fluxuations), and use an
equation which excludes the electroweak interaction and does not include any
mechanism for the production of quark-antiquark pairs. Hence both of these decay
mechanisms are excluded from the theory and thus the mesons described by our
equations cannot decay physically.  Therefore any instability emerging from
these equations will be unphysical, and a sign that the equations are describing
unacceptable states.

\subsection{Unphysical instabilities -- an example}

The Dirac equation for a linear combination of a scalar and
vector confining potential provides a familiar example of the kind of
unphysical instabilities we are discussing. Consider the Dirac
equation for the linear confining potential
$V(r)=\sigma\,r\left\{(1-y)+y\gamma^{0}\right\}$
\begin{equation}
E_{B}\gamma^{0}\phi(r)=(m+V(r)+{\bf \gamma}\cdot{\bf
\nabla})\;\phi(r)\label{intro1}\\
\end{equation}
where $\sigma$ and the vector strength $y$ are both
constants.   The solutions of this equation have both positive and
negative binding energy eigenvalues $E_{B}$.  If the system described by this
equation could interact with the outside world (e.g. absorb or emit photons), 
the positive energy states could decay to negative energy states (unless all
of the negative energy states were occupied as in hole theory).  However, we
have assumed that there is no coupling to the outside world, and hence this
equation should describe a stable system, even if some of the binding energy
eigenvalues are infinitely large and negative.   However, it is well known that
the Dirac equation does not give stable solutions for all values of the
vector strength $y$ and we  review this result now. 

%
\begin{figure}
\begin{center}
\mbox{
   \epsfysize=1.5in
\epsfbox{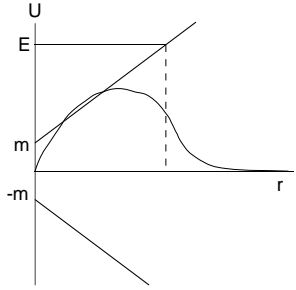}
}
\end{center}
\caption{Sketch of the solution to the Dirac equation for the scalar case,
where $\sigma>0$ and $y=0$.}
\label{figc}
\end{figure}
%
\begin{figure}
\begin{center}
\mbox{
   \epsfysize=1.5in
\epsfbox{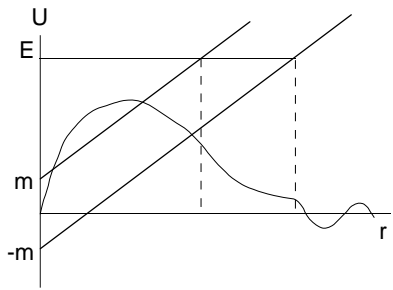}
}
\end{center}
\caption{Sketch of the solution to the Dirac equation for the vector case,
where $\sigma>0$ and $y=1$.}
\label{figd}
\end{figure}

The nature of the solutions to the Dirac equation can be studied by looking 
at the expectation value of $U=m+V$.  The form of this expectation value,
which describes how the wave function behaves, is
\begin{equation}
\langle U \rangle_\pm =\cases{m+\sigma\,r & positive energy \cr
-m -\sigma\,r(1-2y) \qquad & negative energy}
\label{intro2}
\end{equation}
where the positive energy expectation value is a matrix element involving
$u$-type  positive energy spinors, $\langle U \rangle_+ = \bar{u} U u$,
and the negative energy expectation value is a matrix element involving
$v$-type negative energy spinors, $\langle U \rangle_- = \bar{v} U
v$.  The result (\ref{intro2}) comes from the matrix elements
\begin{eqnarray}
\bar{u}u=&1&=-\bar{v}v\nonumber\\
\bar{u}\gamma^{0}u=&1&=\bar{v}\gamma^{0}v\, ,
\end{eqnarray}
which hold when the total momentum $p=0$.
When Eq.~(\ref{intro2}) is
sketched for pure scalar ($y=0$) or pure vector ($y=1$) cases, Fig.~\ref{figc}
and  Fig.~\ref{figd} are produced, respectively.  The resulting wave functions
for a particle with energy $E$ are sketched on the
figures, along with the form of $\langle U \rangle$ which produces it. 

To understand these results, first neglect the coupling between
positive and negative energy states.  Then the positive energy states move
under the influence of the potential $\langle U \rangle_+$ and the negative
energy states under the influence of $\langle U \rangle_-$.   For the scalar
case ($y=0$), the choice $\sigma>0$ produces confinement for both positive and
negative energy states.  Coupling the two solutions does not change this
picture significantly, and the exact solution is a total wave function
which drops to zero at large distances.  This means that both positive and
negative energy solutions describe particles permanently  confined around the
point
$r=0$.  

Next look at the vector case ($y=1$), and begin again by neglecting the
coupling between the positive and negative energy states.  In this case,
however, either the positive or negative energy state is always unconfined.
For the example shown in Fig.~\ref{figd}, $\sigma>0$ and the positive energy
states are confined and the negative energy states are not.  Including
coupling between the positive and negative energy states mixes the two states,
and the wave function for the exact positive energy solution acquires a
component with a ``tail'' which oscillates to infinity, signaling
deconfinement.  The effect of the coupling is to produce an effective potential
composed of two regions separated by a finite  potential barrier
through which the quark can tunnel.  Once it is free of the potential barrier 
it can propagate endlessly through space, thus becoming a free quark. In this
case, the exact coupled solutions do not confine either the positive or
negative energy  states, and the bound state is  unstable. 
This example, known as the Klein paradox \cite{bj}, is one of the unphysical
instabilities we are trying to avoid.
\subsection{Requirements for stability}

A relativisitic equation with a confining kernel with a given Lorentz
structure will have stable, physical solutions only if the following four
conditions are satisfied:
 
\vspace*{0.1in}

\hang (1) the binding energy must be real;
\vspace*{0.1in}

\hang (2) the energy eigenvalues must be independent of the numerical
approximations used to obtain them; 
\vspace*{0.1in}

\hang (3) unphysical solutions, if there are any, must be confined to an
identifiable part of the spectrum clearly separated from the physical
solutions; and
\vspace*{0.1in}

\hang (4) the solutions must have the correct structure in coordinate or
momentum space.
\vspace*{0.1in}

\noindent We will discuss each of these conditions in turn.

{\it Condition\/} 1 {\it -- real energies\/}. Any eigenstate wave
function which describes a meson in momentum space, $\psi({\bf p},t)$, can be
written 
\begin{equation}
\psi({\bf p},t)=\phi({\bf p})e^{-iEt}\,  ,
\label{thr1}
\end{equation}
where $E=\sqrt{\mu^{2}+{\bf P}\,^{2}}$.   The discussion is simplified if the
particle  is chosen to be at rest, ${\bf P}=0$.  Then, if $\mu$ is complex,
$\mu=\mu_{0}\pm i\Gamma/2$, the absolute square of the meson wave function is 
\begin{equation}
|\psi(p,t)|^{2}=|\phi(p)|^{2}e^{\pm\Gamma t}\,  .
\label{thr2}
\end{equation}
As time increases, this goes exponentially either to zero or to infinity,
showing that the state is unstable.  

{\it Condition\/} 2 {\it -- numerical stability\/}.  The different
relativistic equations will be solved numerically in Sec.~IV using
spline functions to model the wavefunctions in momentum space.  (A
description of the properties of the spline functions is given in
Appendix A.)  So long as enough spline functions are used to model the
system, the energy of the lower lying stable states will not vary much
as the spline rank is increased.  However, if the state is unstable it
is part of a continuous spectrum and the energies obtained from the
``eigenvalue'' equation only represent a discrete approximation to
this continuous spectrum. They will vary strongly with the number of
splines, much as the location of the $n$th point in the interval
$[0,1]$ will vary strongly with the number of intervals $N$ into which the
the line segment is divided.  This dependence of an energy level
on spline rank is one of the most obvious symptoms of instability.   

{\it Condition\/} 3 {\it -- isolation of instabilities\/}. In some cases
we find that, following the second criteria, the positive energy
states are stable and the negative energy states are unstable.  This
may be acceptable for a phenomenology, where the negative energy states
can be rejected as unphysical from the start.  However, in some cases
these unstable negative energy states become positive as the spline
number increases, and they can become so positive that they cross the
gap separating the negative and positive energy states, enter the
positive energy spectrum, and mix with states which would otherwise be
stable.  In this case the distinction between (stable) positive energy
states and (unstable) negative energy states becomes blurred, and we
cannot rely on the predictions of the equation.

{\it Condition\/} 4 {\it -- correct structure\/}.  Even if the mass is
real, the state might not be confined in a finite region of coordinate
space (as in the Dirac example outlined above).  If the state is
confined, its coordinate space wave function will approach zero as
$r\to\infty$ {\it faster than an exponential\/}.  It can be shown that
the momentum space wave function resulting from such a state will also
fall off at $p\to\infty$ faster than an exponential, and that the
number of nodes will correspond to the level of the state.  It is easy
to distinguish such behavior from that of an unconfined state, which
is neither localized in coordinate nor momentum space, and which has
many nodes not related to the level of the state.  We can use the
Dirac wave functions for comparison, since we know that they are
stable for scalar confinement and unstable for vector confinement. 
Examples of both types of states will be given in Sec.~IV.

In the following sections, these stability conditions will sometimes be
referred to by number, as we will see that a successful phenomenology
requires that all of them be satisfied. 

\subsection{Summary and Outline}
In summary, the stability conditions are: (1) the eigenvalues of the system
must be real; (2) the eigenvalues cannot vary with the spline rank; (3) the
positive energy states must always be greater than any unstable
negative energy states; and (4) the wave functions must have the
appropriate structure for that specific state.

%
\begin{figure}
\begin{center}
\mbox{
   \epsfysize=1.5in
\epsfbox{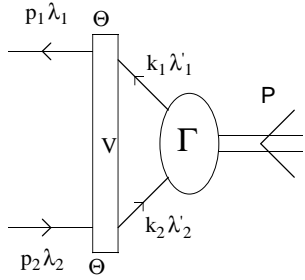}
}
\end{center}
\caption{Feynman diagram for the meson bound state vertex function.  The
kernel, or potential, is denoted by $V$.}
\label{fig15}
\end{figure}

In Sec.~II specific forms are given for the Dirac, Salpeter, and one channel
spectator (denoted 1CS) relativistic equations.  Then in Sec.~III these three
equations are studied using an approximation technique which gives insight
into the origin of the instabilities, and the estimated masses of stable
states are compared to the exact numerical solutions presented in Sec.~IV.
The three equations are solved numerically in Sec.~IV using spline
functions for a quasirelativistic confining potential.  The actual equations 
used in the computer code and the properties of spline functions are given in
Appendix A.  Finally, conclusions are given in Sec.~V.     
\section{The Relativistic Equations}

In this section we define the one-channel spectator (1CS) equation obtained by
confining the heavier particle 1 (assumed to be the quark) to its  positive
energy mass shell, fixing the $k_{0}$ integration. Then we show that these
equations reduce to the Dirac equation for the lighter particle (particle 2) in
the limit when the mass of the heaver particle $m_1\to\infty$. We conclude by
finding a helicity representation for the 1CS and for the Salpeter equation.

\subsection{Dirac form for the one-channel spectator equation}

The Feynman diagram for the bound state meson vertex is shown in
Fig.~\ref{fig15}.  Particle 1 is the quark, particle 2 the antiquark,
and $\Theta$ is a matrix in Dirac space which describes how the confining force
couples to the quark or antiquark.   It can be a scalar,
$\openone$, or the time component of a four-vector, $\gamma^{0}$.  The kernel $V$
contains the momentum dependent structure of the confining potential. The
equations are derived  in the center of mass rest frame, $P=(\mu,{\bf 0})$.  Later,
the quark will be placed on shell, thus producing the single channel equation.
The four momenta used in the diagram are
\begin{eqnarray}
p_{1}=&p+\textstyle{\frac{1}{2}}P   \qquad\; 	
p_{2}&=p-\textstyle{\frac{1}{2}}P\nonumber\\  k_{1}=&k+\textstyle{\frac{1}{2}}P  
\qquad\;		k_{2}&=k-\textstyle{\frac{1}{2}}P \, .\label{der7}
\end{eqnarray}
The vector $k$ is the average internal momentum  
and vector $p$ is the average external momentum of the quark-antiquark pair:
\begin{equation}
p=\textstyle{\frac{1}{2}}(p_{1}+p_{2})   \qquad   P=p_{1}-p_{2}\, .
\end{equation}
With this notation, the Bethe-Salpeter equation \cite{BSp} for
the bound state vertex  function for the meson is 
\begin{equation}
\Gamma (p)=i\int\frac{d^{4}k}{(2\pi)^{4}}\;V(p,k)\;\Theta\;\frac{m_{1}+\not{k}_{1}}
{m_{1}^{2}-k_{1}^{2}}\;\Gamma(k)\;\frac{m_{2}+\not{k}_{2}}{m_{2}^{2}-k_{2}^{2}}
\;\Theta \, .
\label{der10b}
\end{equation}

%
\begin{figure}[t]
\begin{center}
\mbox{
   \epsfysize=1.5in
\epsfbox{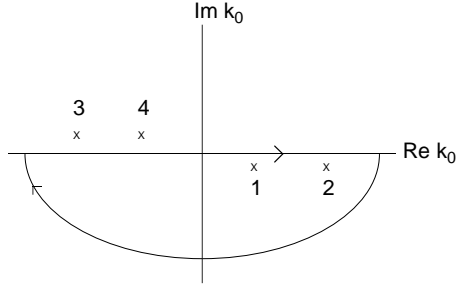}
}
\end{center}
\caption{This figure shows the position of the four poles associated with the
four propagators $G^\rho_i$ in the bound state equations.}
\label{fig16}
\end{figure}

The two fermion propagators have poles in the complex $k_0$ plane; these four
poles are shown Fig.~\ref{fig16}.  Factoring the denominators of the
propagators 
\begin{equation}
\frac{1}
{m_{i}^{2}-k_{i}^{2}}= G_i^+\,G_i^- 
\label{der10a}
\end{equation}
the poles are at
\begin{equation}
\begin{array}{lll} 
{\rm pole\;\;1}\qquad & (G^{+}_{1})^{-1}=E_{k_{1}}-(k_{0}+\frac{1}{2}\mu)-i\epsilon
    =0  & k_{0}=E_{k_{1}}-\frac{1}{2}\mu-i\epsilon \cr
{\rm pole\;\;3} \qquad
&(G^{-}_{1})^{-1}=E_{k_{1}}+(k_{0}+\frac{1}{2}\mu)-i\epsilon
    =0 \qquad & k_{0}=-E_{k_{1}}-\frac{1}{2}\mu+i\epsilon\cr
{\rm pole\;\;2} \qquad & 
(G^{+}_{2})^{-1}=E_{k_{2}}-(k_{0}-\frac{1}{2}\mu)-i\epsilon
    =0   & k_{0}=E_{k_{2}}+\frac{1}{2}\mu-i\epsilon\cr
{\rm pole\;\;4} \qquad  &
(G^{-}_{2})^{-1}=E_{k_{2}}+(k_{0}-\frac{1}{2}\mu)-i\epsilon
    =0  & k_{0}=-E_{k_{2}}+\frac{1}{2}\mu+i\epsilon \, . \cr
\end{array}
\label{der14}
\end{equation}
To place particle 1 on the positive energy mass shell, the $k_0$ integration is
closed in the lower half plane and only the residue from pole 1 is kept.  
This gives the following equation:
\begin{equation}
\Gamma(p)=-\int\frac{d^{3}k}{2E_{k_1}(2\pi)^{3}}\;V(p,k)\;
\Theta\;(m_{1}+\not{\hat k}_{1})
\;\Gamma(k)\;\frac{m_{2}+\not{k}_{2}} {m_{2}^{2}-k_{2}^{2}}\;\Theta \, ,
\label{der10}
\end{equation}
where now $\hat k_1=(E_{k_1}, {\bf k})$ and $k_2=(E_{k_1}-\mu, {\bf k})$. 
This is one form of the one channel spectator equation.

Next, we recall that the projection operator can be written \cite{text} as a sum over
on-shell $u$ spinors (to be defined below)
\begin{equation}
m_{1}+\not{\hat k}_{1}=\sum_{\lambda'} u({\bf k},\lambda')\overline{u}({\bf
k},\lambda')\, .
\label{der18a}
\end{equation}
Therefore, if we define the relativistic meson
wave function by
\begin{equation}
\Psi(k,\lambda) =
\frac{1}{\sqrt{2E_{k_1}}}\;\overline{u}({\bf k},\lambda)
\;\Gamma(k)\;\frac{m_{2}+\not{k}_{2}} {m_{2}^{2}-k_{2}^{2}}
\label{der17a}
\end{equation}
then Eq.~(\ref{der10}) becomes
\begin{equation}
\Psi(p,\lambda) \,(m_{2}-\not{p_{2}})=-\int\frac{d^{3}k}{(2\pi)^{3}}\;
\frac{V(p,k)}{\sqrt{4E_{p_1}E_{k_1}}}
\;\sum_{\lambda'}\Theta^{++}_{\lambda\lambda'}(p,k)\;\Psi(k,\lambda')\;\Theta \, ,
\label{der11a}
\end{equation}
where
\begin{equation}
\Theta^{++}_{\lambda\lambda'}(p,k)= \overline{u}({\bf p},\lambda)\;\Theta\;
u({\bf k},\lambda')
\, .
\label{der11b}
\end{equation}

Equation (\ref{der11a}) is the Dirac form of the one-channel spectator
equation.  Later we will reduce this equation further, but as written in
(\ref{der11a}) it looks very much like a Dirac equation for the light {\it
antiparticle\/} (particle 2) moving under the influence of a effective
potential which depends on the spin of the heavy quark.  To make this
comparison more familiar, we convert the equation into the usual form by taking
the transpose and multiplying by the Dirac charge conjugation matrix
$C$ (see Ref.~\cite{text}).  This gives
\begin{equation}
(m_{2}+\not{p}_{2})\,\hat{\Psi}(p,\lambda) =-\hat{\Theta}\,\sum_{\lambda'}
\int\frac{d^{3}k}{(2\pi)^{3}}\;
\hat{\Psi}(k,\lambda')\;\frac{V(p,k)}{\sqrt{4E_{p_1}E_{k_1}}}
\;\Theta^{++}_{\lambda\lambda'}(p,k) \, ,
\label{der11c}
\end{equation}
where
\begin{equation}
\hat{\Psi}(p,\lambda) = C\,\Psi^{\rm T}(p,\lambda) \qquad 
\hat{\Theta}= C\Theta^{\rm T} C^{-1} \, .
\label{der11d}
\end{equation}
With the exceptions of the spin dependence of the source, expressed through the factor
$\Theta^{++}_{\lambda\lambda'}(p,k)$, and the fact that the effective potential
does not depend solely on ${\bf q}={\bf p}-{\bf k}$, the difference of the three
momenta, Eq.~(\ref{der11c}) looks like the familiar Dirac equation for a particle with
four momentum equal to $-p_2$ (as expected from the charge conjugate state).

We will now calculate the matrix element $\Theta^{++}$ and show that
Eq.~(\ref{der11c}) does indeed reduce to a Dirac equation in the limit
$m_1\to\infty$.  In spin space the $u$ spinor, as
defined in Ref.~\cite{text}, is 
\begin{equation}
u({\bf p},s)=(E_{p}+m)^{\frac{1}{2}}\left(
\begin{array}{c} \openone\\  \\
\displaystyle{\frac{{\bf \sigma}\cdot{\bf p}}{E_{p}+m}}
\end{array}\right)\chi^{s}\, ,
\label{der1}
\end{equation}
which contains the operator ${\bf \sigma}\cdot{\bf p}$.  It is convenient to work
in helicity space, where
${\bf \sigma}\cdot{\bf p}\,\chi^{\lambda}=2\lambda |{\bf p}|\chi^{\lambda}$. 
The $u$ spinor in helicity space is therefore
\begin{equation}
u^+({\bf p},\lambda_{j})\equiv u({\bf p},\lambda_{j})=N_{p_{j}}\left(\begin{array}{c}
\openone\\ \\ 2\lambda_{j}\tilde{p}_{j}\end{array}\right)\chi^{\lambda_{j}},
\label{der2}
\end{equation}
where we have introduced the notation $u^\rho$ with $\rho$-spin $=+1$ for the
positive energy solutions ($u$)  and $\rho$-spin $=-1$ for the negative energy
solutions ($v$, described below), and 
\begin{equation}
N_{p_{j}}=(E_{p_{j}}+m_{j})^{\frac{1}{2}} \qquad
\tilde{p}_{j}=\frac{|{\bf p}|}{N_{p_{j}}^{2}}\, .\label{der4}
\end{equation}
The index $j$ denotes a quark $(j=1)$ or antiquark $(j=2)$.  
The values of $\tilde{p}$ range from 0 to 1.  The helicity spinors are
defined in Table \ref{table1} for cases when the momentum is along the z 
axis (external quarks), and when the momentum is in the $xz$ plane at an angle
$\theta$ with respect to the  z axis.  We will use a prime to distinguish the
latter from the former.  The $v$ spinor, or negative $\rho$-spin state, used in this
paper is 
\begin{equation}
u^-({\bf p},\lambda_{j})\equiv v(-{\bf
p},\lambda_{j})=N_{p_{j}}\left(\begin{array}{c}  -2\lambda_{j}\tilde{p}_{j}\\  \\
\openone \end{array} \right)\chi^{\lambda_{j}}\, ,
\label{der5}
\end{equation}
and is consistent with that used in Ref.~\cite{GVH}.  It is convenient to use the
helicity representation because helicity is invariant under rotations, and because
the vector operator ${\bf \sigma}\cdot{\bf p}$ is replaced by  scalar
eigenvalues, thus simplifying the algebra.

\begin{table}
\begin{center}
\begin{minipage}{3.5in}
\caption{Helicity spinors}\label{table1}
\begin{tabular}{ll}
external quarks & internal quarks\\
\tableline\\
$\lambda_{i}=\phantom{-}\frac{1}{2}$ \qquad  $\left(
\begin{array}{c}1\\0\end{array}\right)$
& $\lambda^{\prime}_{i}=\phantom{-}\frac{1}{2}$ \qquad 
$\left( \begin{array}{c}\cos\frac{\theta}{2}\\  \\ \sin\frac{\theta}{2}
\end{array}\right)$  \\ & \\
$\lambda_{i}=-\frac{1}{2}$ \qquad 
$\left( \begin{array}{c}0\\1\end{array}\right)$ &
$\lambda^{\prime}_{i}=-\frac{1}{2}$ \qquad
$\left( \begin{array}{c} -\sin\frac{\theta}{2}\\ \\ \cos\frac{\theta}{2}
\end{array}\right)$\\[.2cm] 
\end{tabular}
\end{minipage}
\end{center}
\end{table}

The matrix element $\Theta^{++}_{\lambda\lambda'}(p,k)$ is then 
\begin{equation}
\Theta^{++}_{\lambda\lambda'}(p,k)=N_{p_1}N_{k_1}\Delta_{\lambda\lambda'}
(\theta'\theta)\,\left(1
\mp 4\lambda'\lambda \,\tilde{p}_1\tilde{k}_1\right) \, , \label{der5a} 
\end{equation}
where the upper sign is for the scalar vertex  and the lower sign
for the time-like vector case.  We will assume, for the time being,
that the polar angle of the external quark is $\theta'$ instead of 0 (as it will be
later).  Then
\begin{equation}
\Delta_{\lambda\lambda'}(\theta'\theta)=\delta_{\lambda\lambda'}\,
\cos{\textstyle{1\over2}}(\theta-\theta')
-2\lambda\,\delta_{\lambda,\,-\!\lambda'} \, \sin{\textstyle{1\over2}}(\theta-\theta')
\,.\label{der5b}
\end{equation}
Therefore, forming the two independent linear combinations
\begin{eqnarray}
\Phi^+(p)=&&\hat\Psi\left(p,{\textstyle{1\over2}}\right) 
\cos{\textstyle{1\over2}}\theta' - 
\hat\Psi\left(p,-{\textstyle{1\over2}}\right)
\sin{\textstyle{1\over2}}\theta' \nonumber\\
\Phi^-(p)=&&\hat\Psi\left(p,{\textstyle{1\over2}}\right)
\sin{\textstyle{1\over2}}\theta' + 
\hat\Psi\left(p,-{\textstyle{1\over2}}\right)\cos{\textstyle{1\over2}}\theta' 
\end{eqnarray}
Eq.~(\ref{der11c}) becomes
\begin{eqnarray}
(m_{2}+\not{p}_{2})\,{\Phi}^+(p) =-\hat{\Theta}\,&&
\int\frac{d^{3}k}{(2\pi)^{3}}\;\frac{N_{p_1}N_{k_1}}{\sqrt{4E_{p_1}E_{k_1}}}  
\,V(p,k)\nonumber\\ &&\left\{
{\Phi}^+(k)\left[1\mp \tilde{p}_1\tilde{k}_1\cos(\theta-\theta')\right]
\mp{\Phi}^-(k)\;\tilde{p}_1\tilde{k}_1
\sin(\theta-\theta')\right\}\nonumber\\
(m_{2}+\not{p}_{2})\,{\Phi}^-(p) =-\hat{\Theta}\,&&
\int\frac{d^{3}k}{(2\pi)^{3}}\;\frac{N_{p_1}N_{k_1}}{\sqrt{4E_{p_1}E_{k_1}}}    
V(p,k)\nonumber\\ &&\left\{
{\Phi}^-(k)\left[1\mp \tilde{p}_1\tilde{k}_1\cos(\theta-\theta')\right]
\pm{\Phi}^+(k)\;\tilde{p}_1\tilde{k}_1
\sin(\theta-\theta')\right\} \, .\nonumber\\
&&\label{der11f}
\end{eqnarray}
Hence the interaction depends only on the difference $\theta-\theta'$ and we may set
$\theta'=0$ without loss of generality.
 
For later use we will record here the other $\rho$-spin matrix elements of
$\Theta$
\begin{eqnarray}
\Theta^{--}_{\lambda\lambda'}(p,k)=&& \mp\Theta^{++}_{\lambda\lambda'}(p,k)
\nonumber\\
\Theta^{+-}_{\lambda\lambda'}(p,k)=&&-N_{p_1}N_{k_1}\Delta_{\lambda\lambda'}
(\theta'\theta)\,\left(2\lambda' \,\tilde{k}_1 \pm 2\lambda
\,\tilde{p}_1\right) = \pm \Theta^{-+}_{\lambda\lambda'}(p,k)  \, . \label{der5c} 
\end{eqnarray}

\subsection{Limits of the one-channel spectator equation}

Now we can observe that taking the limit $m_1\to\infty$ gives a Dirac
equation for the light particle. The fixed source for the Dirac equation is a heavy
quark, so the equation will model a $Q\,\bar q$ system, such as a $D$ meson.  As
$m_1\to\infty$, $\tilde{p}_1\to0$ and
$V(p,k)\to V({\bf p}-{\bf k})$ (see below), giving
\begin{eqnarray}
(m_{2}-\not\!{p'}_{2})\,{\Phi}(p) =-\hat{\Theta}\,
\int\frac{d^{3}k}{(2\pi)^{3}}\;V({\bf p}-{\bf k})
{\Phi}(k) \, ,
\label{der11g}
\end{eqnarray}
where the helicity of the heavy particle was dropped because the equation is
independent of it and we introduced the physical momentum
$p_2'=-p_2=(\mu-E_{p_1},-{\bf p})\to (E_B, {\bf p'})$, with $E_B=\mu-m_1$.  In
position space Eq.~(\ref{der11g}) is
\begin{eqnarray}
(m_{2}-E_B\gamma^0 +i\,\gamma^i\partial_i)\,{\Phi}({\bf r}) =-\hat{\Theta}\;V({\bf r})
{\Phi}({\bf r}) \, ,
\label{der11h}
\end{eqnarray}
We will return to this equation in the next subsection.

The confining potential ${V}({\bf r})$ which appears in Dirac Eq.~(\ref{der11h}) is
taken to be a simple linear potential in position space \cite{gm} 
\begin{equation}
V(r)=\sigma r =\lim_{\epsilon\rightarrow 0}\sigma r e^{-\epsilon r}
=\lim_{\epsilon\rightarrow 0}\sigma \frac{d^{2}}{d\epsilon^{2}}
\frac{ e^{-\epsilon r}}{r}. \label{iddr27}
\end{equation}
In momentum space this potential is
\begin{equation}
V({\bf q})=-8\pi\sigma\lim_{\epsilon\rightarrow
0}\left\{\frac{1}{({\bf q}^{2}+\epsilon^{2})^{2}}
-\frac{4\epsilon^{2}}{({\bf q}^{2}+\epsilon^{2})^{3}} \right\}. 
\end{equation}
This form of the potential is inconvenient because the limit $\epsilon\to0$ must be
taken numerically.  For the Dirac equation, we use an alternative form which
has the same physics 
\begin{equation}
V({\bf q})=-{8\pi\sigma}\left\{\frac{1}{{\bf q}^{4}}-\delta^{3}({q})
\int \frac{d^{3}q^{\prime}}{{\bf q}^{\prime 4}} \right\}
\label{iddr28}
\end{equation}
where ${\bf q}={\bf p}-{\bf k}$ and ${\bf q}^{\prime}={\bf p}-{\bf k}^{\prime}$. 
It is instructive to note that the position space form of each of the terms in
(\ref{iddr28}) is
\begin{eqnarray}
-\frac{8\pi\sigma}{{\bf
q}^{4}}&&=\lim_{\epsilon\rightarrow 0} -\frac{8\pi\sigma}{({\bf
q}^{2}+\epsilon^{2})^{2}}\rightarrow \lim_{\epsilon\rightarrow 0} -\sigma\frac
{e^{-\epsilon r}}{\epsilon} \simeq -\frac{\sigma}{\epsilon} + \sigma r + \cdots 
\nonumber\\
\delta^{3}({q})\int d^{3}q^{\prime} \frac{8\pi\sigma}{{\bf q}^{\prime 4}} &&
=\lim_{\epsilon\rightarrow 0} \delta^{3}({q})\int d^{3}q^{\prime}
\frac{8\pi\sigma} {({\bf
q}^{2}+\epsilon^{2})^{2}} \rightarrow \frac{\sigma}{\epsilon}\, . \label{e7}
\end{eqnarray}
Hence the role of the delta function subtraction is to remove the infinite
constant from the first term, leaving a pure linear potential.

In relativistic two-body equations, the potential (\ref{iddr28}) is generalized to
\cite{gm}
\begin{equation}
V(p,k)=-{8\pi\sigma}\left\{\frac{1}{(p-k)^{4}}-E_{p_1}\,\delta^{3}(p-k)
\int \frac{d^{3}k^{\prime}}{E_{k'_1}\,(p-k')^4} \right\}\, .
\label{iddr28rel}
\end{equation}
The insertion of the energy factors is necessary to make the kernel (\ref{iddr28rel})
covariant, and is associated with the restriction of the heavy quark to its
mass-shell\cite{gm}.  The full 1CS also includes the covariant replacement
$(p-k)^2\to (E_{p_1}-E_{k_1})^2-({\bf p}-{\bf k})^2$, but in both the theoretical and
numerical studies in this paper we have neglected retardation and use
the simplest replacement $(p-k)^2\to -({\bf p}-{\bf k})^2$.  We will refer to this
as the quasirelativistic approximation, and it should be emphasized that we use
this approximation in this paper only to simplify the discussion.  We also
neglect the  regularization factor and form factor introduced in previous
studies \cite{gm}.

The energy factor in the subtraction term in (\ref{iddr28rel}) gives rise to a
relativistic effect of some importance.  To see this, evaluate the diagonal matrix
element 
\begin{eqnarray}
\left<\psi|V_{C}|\psi\right>&&=\int d^3p\,d^3k \,\psi(p) \left\{E_{p_1}\,\delta^{3}(p-k)
\int \frac{d^{3}k^{\prime}\, 8\pi\sigma}{E_{k'_1}\,(p-k')^4}\right\} \psi(k)
\nonumber\\
=&&\int \frac{d^3r\, d^3r'}{(2\pi)^3}\, d^3p\, \psi(r') e^{i{\bf p}\cdot({\bf r}'-
{\bf r})}\left\{E_{p_1}
\int \frac{d^{3}k^{\prime}\,
8\pi\sigma}{E_{k'_1}\,(p-k')^4}\right\}\psi(r)\nonumber\\ 
=&&\int d^3r \psi(r) \left\{E_{p_1}
\int \frac{d^{3}k^{\prime}\,
8\pi\sigma}{E_{k'_1}\,(p-k')^4}\right\}\psi(r) \, , 
\end{eqnarray}
where, in the last line, $p=\sqrt{p^2}\to \sqrt{-\nabla^2}$.  Hence the subtraction
term becomes an operator which is a function of $p^2\to -\nabla^2$ but independent of $r$. 
It  can be evaluated by standard means
\begin{eqnarray}
E_{p_1}\int \frac{d^{3}k^{\prime}\,8\pi\sigma}{E_{k'_1}\,(p-k')^4}=&&
\lim_{\epsilon\rightarrow 0} E_{p_1}\int
\frac{d^{3}k^{\prime}\,8\pi\sigma}{E_{k'_1}\,[(p-k')^2+\epsilon^2]^2}\nonumber\\
=&&\frac{\sigma}{\epsilon} -\frac{2\sigma}{\pi m_1}\left\{ \frac{m_1}{E_{p_1}} +
\frac{m_1^3}{p E_{p_1}^2}\log\left(\frac{E_{p_1}+p}{m_1}\right)\right\}\nonumber\\
=&&\frac{\sigma}{\epsilon} -C(p^2)\, .
\label{eq8}
\end{eqnarray}
This new subtraction term contains the same singular part we found before
[see Eq.~(\ref{e7})] {\it plus\/} a new finite part $C(p^2)$.  The finite part arises
from the relativistic energy factor in Eq.~(\ref{eq8}), which produces an
infinitesimal  modification of the singular part, and it vanishes in the
$m_1\to\infty$ limit.  It has an interesting effect which will be discussed in the
next subsection.

The confining potential has the property that it is very singular when ${\bf
q}\to0$.  This suggests using a peaking approximation in which
$\theta\simeq\theta'$, so that the coupling between
$\Phi^+$ and $\Phi^-$ can be neglected.  No further approximations are needed,
because we may reduce all factors of $E_{p_1}$ and $E_{k_1}$ to derivative operators, and
replace
\begin{eqnarray}
\tilde{k}_1\tilde{p}_1 \cos(\theta-\theta')=&& \frac{{\bf k}\cdot {\bf p}}{(E_{k_1}
+m_1)(E_{p_1}+m_1 )}\nonumber\\
\to&& -\frac{({\bf k}- {\bf p})^2}{2(E_{k_1}+m_1)(E_{p_1}+m_1 )}
+\frac{\tilde{k}_1^2}{2} +\frac{\tilde{p}_1^2}{2}\, ,
\end{eqnarray}
because the operators $k^2$ (which operates on the initial wave functions) and
$p^2$ (which operates on the final wave functions) will eventually give indentical
results.  Furthermore, the operator ${\bf q}^2$ can be reduced to
\begin{eqnarray}
\frac{8\pi\sigma{\bf q}^2}{2{\bf q}^4}=\frac{4\pi\sigma}{{\bf
q}^2}\to\frac{\sigma}{r} \, .
\end{eqnarray}

Combining all of these effects, Eq.~(\ref{der11f}) becomes a single
Dirac-like equation with a momentum dependent potential 
\begin{eqnarray}
(m_{2}+\not\!{p}_{2})\,{\Phi}(p) =&&\hat{\Theta}C(p)\Phi(p)
-\hat{\Theta}\,\int\frac{d^{3}k}{(2\pi)^{3}}\;V(p,k) {\cal N}
{\Phi}(k)\nonumber\\
&&\times\left[1\mp\frac{1}{2}(\tilde{k}_1^2+\tilde{p}_1^2) \pm \frac{1}{2}
\frac{{\bf q}^2}{(E_{k_1}+m_1)(E_{p_1}+m_1 )} \right] \, ,
\label{der11i}
\end{eqnarray}
where ${\cal N}={N_{p_1}N_{k_1}}/\sqrt{4E_{p_1}E_{k_1}}$.
Recalling that $p_{20}=M-E_{p_1}=E_B+m_1-E_{p_1}$, the coordinate space form of this
equation is
\begin{eqnarray}
&&\left(m_{2}-\left[E_B+m_1-\sqrt{m_1^2-\nabla^2}\right]\;\gamma^0
+i\,\gamma^i\partial_i\right)\,{\Phi}({\bf r})\nonumber\\
 &&\qquad=  -\hat{\Theta}\left(\left[\sigma r - C(p^2)\right]\left(1\mp
\tilde{p}^2_1\right)\mp \frac{\sigma}{r}\frac{1}{(E_{1}+m_1)^2}\right) {\cal N}{\Phi}({\bf r})
\, ,
\label{der11j}
\end{eqnarray}
where we have anticipated the application to {\it diagonal\/} matrix elements where
$k^2=p^2\to-\nabla^2$, and all functions $E_{p_1}=E_1$ are replaced by
$\sqrt{m_1^2-\nabla^2}$.  Using Eq.~(\ref{der11j}) we can study the single-channel spectator
equation when the mass $m_1$ is close to $m_2$, and also see how it approaches the Dirac
limit.  We will study these issues approximately in Sec.~III.

\subsection{Equations with mixed scalar and vector confinement} 

Note that the operator ${\Theta}$ depends on its Dirac
structure; it is $\openone $ for a scalar confinement and
$\gamma^0$ for vector confinement.  Hence
Eq.~(\ref{der11d}) gives
\begin{eqnarray}
\hat{\Theta}= \cases{ \phantom{-}\openone &\qquad{\rm scalar}\cr
-\gamma^0&\qquad{\rm vector}\, .}\label{t1}
\end{eqnarray}
In the nonrelativistic limit, the Dirac equation reduces to a Schr\"odinger equation
for the upper component, and we will choose the sign of our potential so
that it confines the positive energy solution in the nonrelativistic limit. Hence, in order
to obtain a nonrelativistic confining potential equal to $\sigma r$, independent of the mixing
parameter $y$, the operator form ${\cal O}$ of a mixed kernel must be
\begin{eqnarray}
{\cal O}= (1-y)\,\openone\,\otimes\,\openone -y\,\gamma^0\,\otimes\,\gamma^0 \, .
\label{t2}
\end{eqnarray}
Using this definition and the result  Eq.~(\ref{der11j}) gives the following equation for a
mixed confining potential
\begin{eqnarray}
&&\left(m_{2}-\left[E_B+m_1-E_1\right]\;\gamma^0
+i\,\gamma^i\partial_i\right)\,{\Phi}({\bf r})\nonumber\\
 &&\qquad=  -\hat{\Theta}\left(\left[\sigma r - C(p^2)\right]\left(1-
\tilde{p}^2_1(1-2y)\right)-\frac{\sigma}{r}\frac{(1-2y)}{(E_{1}+m_1)^2}\right)
{\cal N} {\Phi}({\bf r}) \, .
\label{der11l}
\end{eqnarray}
Assuming a ground state solution of the form \cite{text}
\begin{eqnarray}
{\Phi}({\bf r})=\left(\matrix{ f(r) \cr -ig(r)\sigma\cdot{\bf\hat{r}}}\right)\chi
\, , \label{der11m}
\end{eqnarray}
Eq.~(\ref{der11l}) reduces to the following set of coupled equations for the radial
wave functions $f(r)$ and $g(r)$
\begin{eqnarray}
&&\left(E_B-m_{2}-\left[E_1-m_1\right] \right)\,f \,+\,\frac{dg}{dr}+\frac{2}{r}g\nonumber\\
&&\qquad=\left((\sigma r-C) \left[1-
\tilde{p}^2_1\,(1-2y)\right] -\frac{\sigma}{r} \frac{(1-2y)}{(E_{1}+m_1)^2}\right) 
{\cal N}\,f \nonumber\\
&&\left(E_B + m_{2}-\left[E_1-m_1\right]\right)\,g\, -\, \frac{df}{dr}\nonumber\\
&&\qquad=-\left((\sigma r-C)
\left[(1-2y) -\tilde{p}^2_1\right] - \frac{\sigma}{r}
\frac{1}{(E_{1}+m_1)^2}\right) {\cal N}\,g
\, .
\label{der11n}
\end{eqnarray}
We will return to this coupled set of equations in the next section.

\subsection{Spectator equation in helicity space}

The equations we have obtained so far are convenient for approximate analysis, but
an exact helicity decomposition is better for numerical solutions.  To
obtain this form of the one-channel spectator equation, return to Eq.~(\ref{der11a})
and expand the wave function and the projection operator in terms of the helicity
spinors given in Eqs.~(\ref{der2}) and (\ref{der5}).  The wave function can be
expanded using the decomposition of the propagator into $\rho$ spin 
contributions \cite{text}, 
\begin{equation}
\frac{(m_{2}+ \not{k}_{2})}{m_{2}^{2}-k_{2}^{2}}=\frac{1}{2E_{k_2}}
\sum_{\lambda_2}
\left[\frac{u({\bf k},\lambda_2)\bar{u}({\bf k},\lambda_2)}  
{E_{k_{2}}-k_{20}-i\epsilon}-
\frac{v(-{\bf k},\lambda_2)\bar{v}(-{\bf k},\lambda_2)}
{E_{k_{2}}+k_{20}-i\epsilon}\right]\, .
\label{der9}
\end{equation}
Using this in Eq.~(\ref{der17a}) shows that the wave function has the form
\begin{eqnarray}
\Psi(p,\lambda)=\sum_{\rho\lambda_2}\Psi^\rho_{\lambda\lambda_2}(p)\,
\overline{u}^\rho({\bf p},\lambda_2) \, .
\label{der9a}
\end{eqnarray}
Furthermore, the most general form of the pseudoscalar vertex function with
particle 1 on shell is
\begin{equation}
\overline{u}^+({\bf p},\lambda)\,\Gamma(p)=\overline{u}^+({\bf p},\lambda)\;
\left\{\Gamma_1\,\gamma^{5}+\Gamma_2\,
\gamma^{5}\,\left(m_{2}-\not\!{p}_{2}\right)\right\}\, ,
\label{der19}
\end{equation}
and these Dirac operators are built only from the $2\times2$ matrices $\openone$ and
$\sigma\cdot {\bf p}=2\lambda p$.   Therefore in helicity space the helicity is
conserved and an explicit calculation shows that
\begin{eqnarray}
\overline{u}^+({\bf p},\lambda)\,\Gamma(p)\,u^\rho({\bf
p},\lambda_2)=\delta_{\lambda\lambda_2}\,(2\lambda)^{\delta_{+\rho}}\,\Gamma^\rho(p)
\, ,
\label{der11o}
\end{eqnarray}
where the $\Gamma^\rho(p)$ are independent of the helicity.  Hence the expansion
(\ref{der9a}) can be written
\begin{eqnarray}
\sqrt{2E_{p_2}}\,\Psi(p,\lambda)=\frac{1}{\sqrt{4E_{p_2}E_{p_1}}}\left\{
\psi_{1a}(p)\,\overline{u}^-({\bf p},\lambda) + (2\lambda)\,
\psi_{1b}(p)\,\overline{u}^+({\bf p},\lambda) \right\}\, ,
\label{der20aa}
\end{eqnarray}
where
\begin{eqnarray}
\psi_{1a}=-\frac{\Gamma^{-}}{E_{k_{2}}+E_{k_{1}}-\mu} & \qquad &
\psi_{1b}=\frac{\Gamma^{+}}{E_{k_{2}}-E_{k_{1}}+\mu},
\label{der22}
\end{eqnarray}
Bringing all of these elements together, using Eq.~(\ref{der5c}), and choosing the
sign of the vector interaction in accordance with Eq.~(\ref{t2}) gives the helicity
form of the single channel spectator equation 
\begin{equation}
\left(\begin{array}{c}(E_B-E_{p_2}-[E_{p_1}-m_1])\,\psi_{1a}(p)\\ 
(E_B +E_{p_2}-[E_{p_1}-m_1])\,\psi_{1b}(p)\\ 
\end{array}\right)=\int_{\bf{k}} \overline V\,\pmatrix{D_{1} & D_{2} \cr
D_{3} & D_{4} \cr}
\left(\begin{array}{c}\psi_{1a}(k)\\ \psi_{1b}(k)\\ 
\end{array}\right)\, ,
\label{der23}
\end{equation}
where $E_B=\mu-m_1$,
\begin{equation}
\int_{\bf {k}}=\int\frac{d^3k}{(2\pi)^3} \; ,
\label{der23a}
\end{equation}
the rescaled potential kernel is
\begin{equation}
\overline V=\frac{N_{p_1}N_{p_2}N_{k_1}N_{k_2}} 
{4\sqrt{E_{p_1}E_{p_2}E_{k_1}E_{k_2}}}\;V \, ,
\label{der25a}
\end{equation} 
and $D_i=A_i + B_i \cos\theta$ with 
\begin{equation}
\begin{array}{ll}
A_{1}=Q \qquad & B_{1}=\mp R \cr
A_{2}=T_{2} \qquad & B_{2}=\pm S_{2} \cr
A_{3}=S_{2} \qquad & B_{3}= \pm T_{2} \cr
A_{4}= R \qquad & B_{4}= \mp Q \, , \label{der25}
\end{array} 
\end{equation}
and
\begin{equation}
\begin{array}{ll}
Q=1+\tilde{p}_1\tilde{p}_2\tilde{k}_1\tilde{k}_2 \qquad & 
R=\tilde{p}_{1}\tilde{k}_{1}+\tilde{p}_{2}\tilde{k}_{2}\cr
S_{j}=\tilde{p}_{j}-\tilde{k}_1\tilde{k}_2 \tilde{p}_{j'} \qquad &  
T_{j}=\tilde{k}_{j}- \tilde{p}_1\tilde{p}_2 \tilde{k}_{j'} \, .\label{der25c}
\end{array}
\end{equation}
In (\ref{der25}) the upper sign holds for scalar confinement and the lower for vector
confinement and in (\ref{der25c}) $j'\ne j$.

For the mixed scalar/vector confinement defined in Eq.~(\ref{t2}) the values of
$A_i$ and $B_i$ are:
\begin{equation}
\begin{array}{ll}
A_{1}= Q \qquad & B_{1}=-R(1-2y) \cr
A_{2}= T_{2} \qquad & B_{2}=S_{2}(1-2y) \cr
A_{3}=S_{2} \qquad & B_{3}=T_{2}(1-2y) \cr
A_{4}= R \qquad & B_{4}=-Q(1-2y) \, .\label{der25b}
\end{array}
\end{equation}
When the masses are equal this equation reduces to the equation
previously introduced in Ref.~\cite{gm2}.

\subsection{Dirac equation in helicity space}

The helicity form of the Dirac equation is obtained from (\ref{der23}) by taking the
$m_1\to\infty$ limit 
\begin{equation}
\left(\begin{array}{c}(E_B-E_{p_2})\,\psi_{1a}(p)\\ 
(E_B +E_{p_2})\,\psi_{1b}(p)\\ 
\end{array}\right)=\int_{\bf{k}} \bar V\,\pmatrix{d_{1} & d_{2} \cr
d_{3} & d_{4} \cr}
\left(\begin{array}{c}\psi_{1a}(k)\\ \psi_{1b}(k)\\ 
\end{array}\right)\, ,
\label{dirac1}
\end{equation}
where now $\bar V=V N_{p_2}N_{k_2}/\left(2\sqrt{E_{p_2}E_{k_2}}\,\right)$ and 
$d_i=a_i + b_i
\cos\theta$ with
\begin{equation}
\begin{array}{ll}
a_{1}=1 &b_{1}=-\tilde{p}_2\tilde{k}_2(1-2y)  \cr
a_{2}=\tilde{k}_2 &b_{3}=\tilde{p}_2(1-2y)   \cr
a_{3}=\tilde{p}_2 &b_{2}=\tilde{k}_2(1-2y)  \cr
a_{4}=\tilde{p}_2\tilde{k}_2\qquad &b_{1}=-(1-2y)  \, .  
\label{iddr26a}
\end{array}
\end{equation}

We conclude this section with a derivation of the helicity form of the Salpeter
equation.

\subsection{Salpeter equation in helicity space}

The Salpeter \cite{SpE} uses the approximation that the potential, or kernel, of the
Bethe-Salpeter equation is independent of $k_{0}$ and $p_{0}$. 
Therefore, in coordinate space  the potentials and the wave functions are
instantaneous, i.e. $t_{1}=t_{2}$.  The Salpeter Equation has two undesirable
features. First, neglecting the energy dependence of the
kernel is unphysical.   Second, there is no Dirac limit 
for this equation.  When the mass of one of the particles is taken to infinity,
the resulting equations do not  reduce to a Dirac equation for the light quark
moving in the field created by the heavy quark.  Hence, it is most appropriate to use this
equation for equal masses, far away from the one body limit.

The direct derivation of the Salpeter equation utilizes the same steps as those
used for the 1CS equation with a few modifications.  In this case pole 2, as defined in
Eq.~(\ref{der14}) and  Fig.~\ref{fig16}, must be included.   For brevity we will only
give the final result.  The general equation is
\begin{equation}
\Gamma^{\rho_{1}\rho_{2}}_{1}(p)=\int\frac{d^{3}k}{(2\pi)^{3}}V_{11}^{\prime}
\sum_{\rho_{1}^{\prime}\rho_{2}^{\prime}}\sum_{\lambda_{1}^{\prime}
\lambda_{2}^{\prime}}
\Theta_{1}^{\rho_{1}\rho_{1}^{\prime}}\Gamma_{1}^{\rho_{1}^{\prime}\rho_{2}^{\prime}}
G_{2}^{\rho_{2}^{\prime}}
\Theta_{2}^{\rho_{2}^{\prime}\rho_{2}}
\label{spder1}
\end{equation}
where $\rho_{1}\neq\rho_{2}$ and $\rho_{1}^{\prime}\neq\rho_{2}^{\prime}$. 
The second channel wave function, denoted $\psi_{2a}$, corresponds to propagation of the two
quarks in their negative energy state, and is equal to 
\begin{equation}
\psi_{2a}=-\frac{\Gamma^{-+}_{1}}{E_{k_{2}}+E_{k_{1}}+\mu}\, .
\label{spder2}
\end{equation}
The two wave functions, $\psi_{1a}$ and $\psi_{2a}$ satisfy  the coupled
equations
\begin{equation}
\left(\begin{array}{c}(\mu-E_{p_2}-E_{p_1})\psi_{1a}(p)\\ 
(\mu+E_{p_2}+E_{p_1})\psi_{2a}(p)\\ 
\end{array}\right)=\int_{\vec{k}} \overline{V}\,\pmatrix{D_{1} & -D_{5} \cr
D_{5} & -D_{1} \cr}
\left(\begin{array}{c}\psi_{1a}(k)\\ \psi_{2a}(k)\\ 
\end{array}\right)\, ,
\label{spder2a}
\end{equation}
with
\begin{equation}
D_{5}=\tilde{p}_{1}\tilde{p}_{2}+\tilde{k}_{1}\tilde{k}_{2}
+ (1-2y)\left(\tilde{p}_{1}\tilde{k}_{2}+\tilde{p}_{2}\tilde{k}_{1}\right) \cos\theta 
\label{spder3}
\end{equation}
All other terms have the same definitions as before.

\section{Approximate theoretical results}

In this section we develop approximations which help us 
understand the stability issues which will arise when the equations are solved numerically. 

\subsection{Dirac solutions for large $r$}

We begin by studying the stability of Eq.~(\ref{der11n}) in the Dirac limit when
$m_1\to\infty$
\begin{eqnarray}
&&E_B\,f=\left(m_{2}+\sigma r \right)\,f -\frac{dg}{dr}-\frac{2}{r}g
\nonumber\\
&&E_B\,g=\left(-m_{2} -\sigma r (1-2y) \right)\,g +\frac{df}{dr}\, .
\label{der11p}
\end{eqnarray}
This is the exact Dirac equation for a potential which is a
superposition of scalar and vector linear confining forces.  At large $r$
the equations become approximately
\begin{eqnarray}
\sigma r \,f -\frac{dg}{dr}&&=0
\nonumber\\
 -\sigma r (1-2y) \,g +\frac{df}{dr}&&=0\, .
\label{der11q}
\end{eqnarray}
The solution to these equations depends on the value of $y$.  If 
$y<1/2$, then the solution which approaches zero as $r\to\infty$ is
\begin{eqnarray}
f(r)=N_f\,e^{-\sqrt{1-2y}\;\frac{1}{2}\sigma r^2} \qquad g(r)=
N_g\,e^{-\sqrt{1-2y}\;\frac{1}{2} \sigma r^2}\, ,
\label{r1}
\end{eqnarray}
where 
\begin{eqnarray}
N_f=-\sqrt{1-2y}\;N_g \, .
\label{r2}
\end{eqnarray}
Note that the wave functions become less confined as $y\to1/2$.  For $y>1/2$ the
solutions are oscillatory and escape to large $r$.  In this case the most general solution 
is a linear combination of the following two independent solutions 
\begin{eqnarray}
f_1(r)=&&N_{f_1}\,\sin\left(\sqrt{2y-1}\;\frac{1}{2}\sigma r^2\right) \qquad g_1(r)=
N_{g_1}\,\cos\left(\sqrt{2y-1}\;\frac{1}{2}\sigma r^2\right) \nonumber\\
f_2(r)=&&N_{f_2}\,\cos\left(\sqrt{2y-1}\;\frac{1}{2}\sigma r^2\right) \qquad g_2(r)=
N_{g_2}\,\sin\left(\sqrt{2y-1}\;\frac{1}{2}\sigma r^2\right) \, ,
\label{r3}
\end{eqnarray}
where
\begin{eqnarray}
N_{f_1}&&=-\sqrt{2y-1}\;N_{g_1} \nonumber\\
N_{f_2}&&=\sqrt{2y-1}\;N_{g_2} \, .
\label{r4}
\end{eqnarray}
This is the simple mathematical explanation behind the results shown in
Figs.~\ref{figc} and \ref{figd}.

\subsection{Estimates for the one-channel spectator equation}

Study of the solutions of the approximate one channel spectator equation,
 Eq.~(\ref{der11n}), is complicated by
the presence of the operators $\sqrt{m_1^2-\nabla^2}$ and $\tilde{p}_1^2$. 
We will therefore develop a variational-like method which can give us
insight into the confining behavior of the equation.

First, if $\sigma=0$ the {\it exact\/} solution of the
equations is
\begin{eqnarray}
f(r)=&&f_o\,j_0(\gamma r) \nonumber\\
g(r)=&&g_o\,j_1(\gamma r) \, , 
\label{r5a}
\end{eqnarray}
where $j_\ell$ is the spherical Bessel function of order $\ell$, and the energy is a
function of the parameter $\gamma$ 
\begin{eqnarray}
E_B(\gamma) =\left[\sqrt{m_1^2+\gamma^2} - m_1\right] \pm \sqrt{m_2^2+\gamma^2}\,  .
\label{sol1}
\end{eqnarray}
The spectrum is continous with a gap between the positive and negative energy states.  It 
is amusing to see that the energies of both the positive and negative energy states are
always {\it greater\/} than the corresponding Dirac state energies, and that the negative
energy spectrum is now bounded between $-m_2$ and $-m_1$, instead of running from $-m_2$ 
to $-\infty$.  This already illustrates one of the new features of the 1CS equation.

\begin{figure}[t]
\vspace*{-0.2in}
\mbox{
\leftline{\hspace*{-1.0in}
   \epsfysize=3.0in
\epsfbox{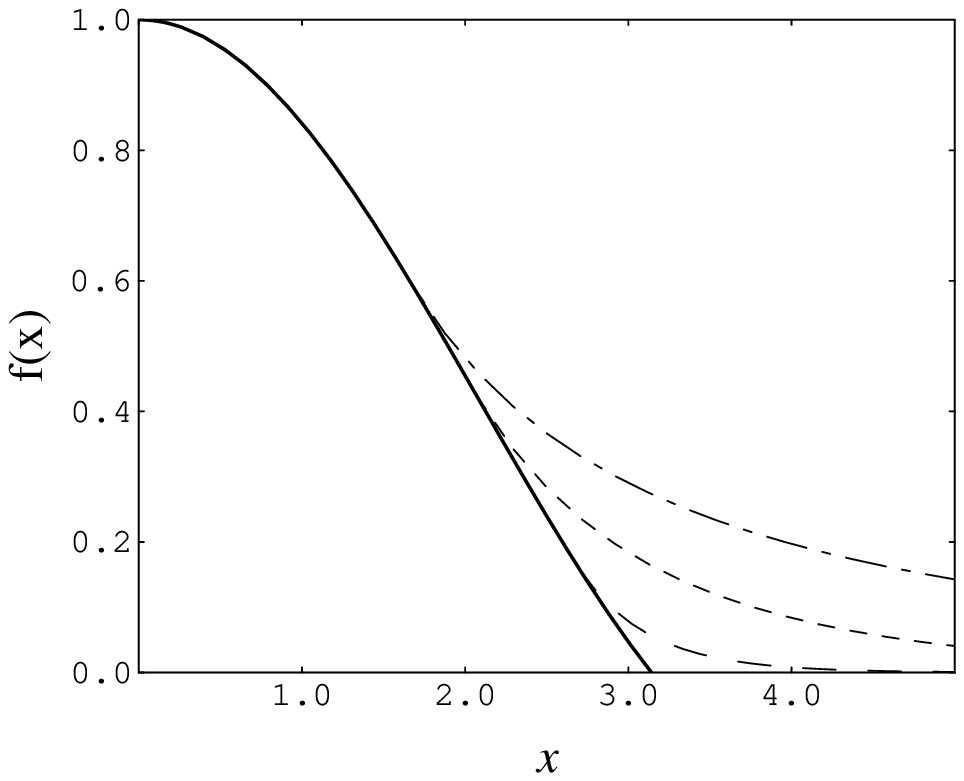}}
\vspace*{-3.0in}
\leftline{\hspace*{-4.2in}
   \epsfysize=3.0in
\epsfbox{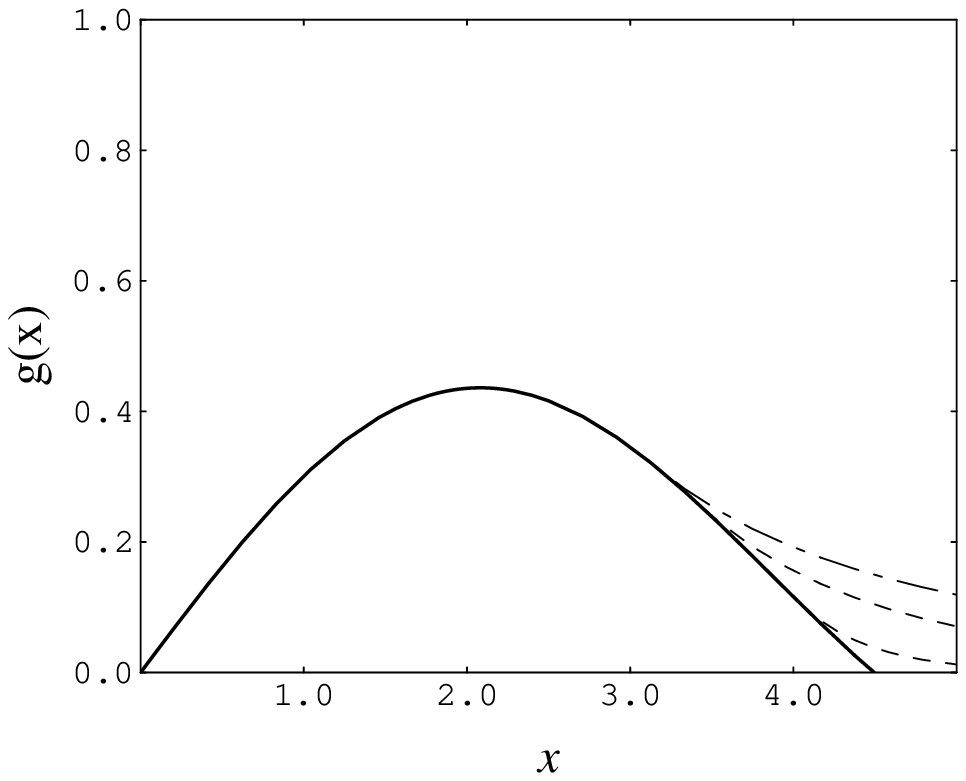}}}
\caption{The trial wave functions $f(x)$ and $g(x)$ as a
function of $x=\gamma r$.  The different tails are for the
cases $k/\gamma$ = 0.1 (biggest tail), 0.5, and 2.0
(smallest tail), as discussed in the text.}
\label{wave}
\end{figure}

When $\sigma\ne0$ we cannot solve the equation analytically, and will limit our study to 
the behavior of the expectation value of the energies as estimated by taking matrix
elements of the equation.  To compute these matrix elements we will use wave functions of 
the type shown in Fig.~\ref{wave}, which are constructed from spherical bessel functions 
of order zero and one.  This choice makes the evaluation of functions of the
operator $\nabla^2$ easy.

Ideally, the functions used should consist of a region where $-\nabla^2$ is positive, and a
``tail'' region where $-\nabla^2<0$.  The functions shown in Fig.~\ref{wave} were
constructed from $j_\ell(\gamma r)$ and $h_\ell(k r)$ joined so that the function and 
its first derivative are continuous.  However, we found that the
contributions from the tails did not change the qualitative behavior of the
matrix elements, and hence we present here only the simplest results for
wave functions without tails (where $k/\gamma\to\infty$, the heavy solid
lines in the figure).  These results are easy to evaluate.   

%

Hence the ``trial'' wave functions we choose are
\begin{eqnarray}
f(r)=&&\cases{f_o\,j_0(\gamma r)  & $\gamma r<\pi$ \cr
0\qquad\qquad\qquad & $\gamma r>\pi$} \nonumber\\
g(r)=&&\cases{g_o\,j_1(\gamma r)  & $\gamma r< n_1$ \cr
0\qquad\qquad\qquad & $\gamma r> n_1$\, ,} 
\label{r5}
\end{eqnarray}
where $j_\ell$ is the spherical Bessel function of order $\ell$, $\gamma$ is a
variational parameter, and the constant
$n_1= 4.493$ is the location of the zero of $j_1$.  These wave
functions are eigenfunctions of the operator
$\nabla^2$:
\begin{eqnarray}
\nabla^2\,f(r)&&=\frac{1}{r}\frac{\partial^2}{\partial
r^2}\,r\,f(r)=-\gamma^2\;f(r)\nonumber\\ 
\nabla^2\,g(r)&&=\left(\frac{1}{r}\frac{\partial^2}{\partial
r^2}\,r -\frac{2}{r^2}\right)\, g(r)= -\gamma^2\;g(r)\, \, .
\label{r6}
\end{eqnarray}
Hence the operators
$\sqrt{m_1^2-\nabla^2}$ and $\tilde{p}_1^2$ can be readily calculated.

%
\begin{figure}[t]
\mbox{
   \epsfysize=4.0in
\epsfbox{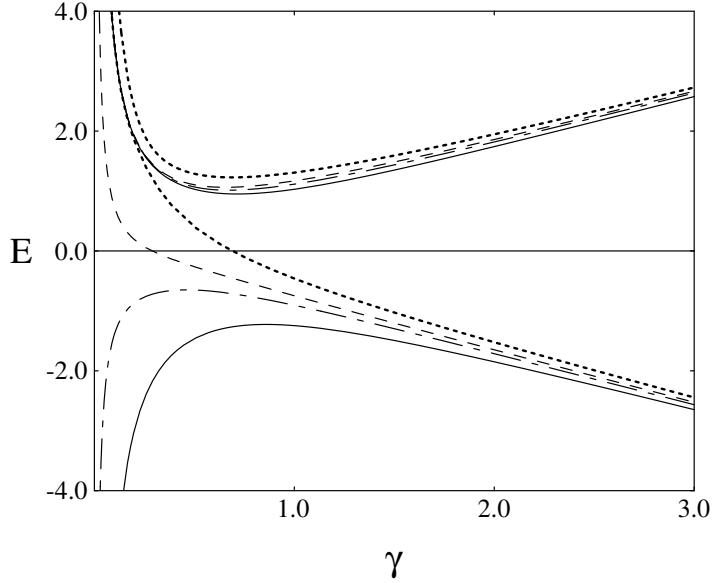}
}
\caption{The Dirac energy $E=E_B$ as a function
of the variational parameter $\gamma$ for different mixing ratios $y=0$ (solid
line), $y=0.4$ (dot-dashed), $y=0.6$ (dashed), and $y=1.0$ (dotted).}
\label{v1}
\end{figure}

%
\begin{figure}
\mbox{
   \epsfysize=3.8in
\epsfbox{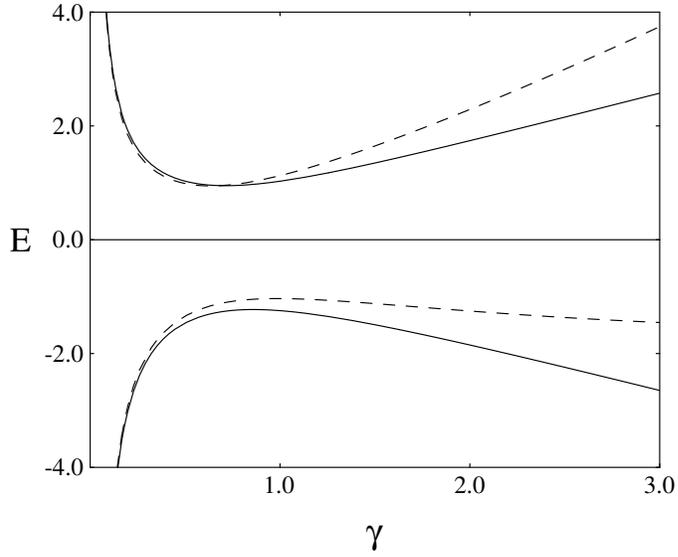}
}
\caption{Energies $E=E_B$ as a function
of the variational parameter $\gamma$ for the Dirac equation (solid line) and the
1CS equation with $m_1=10 m_2$ (dashed line).  In both cases, $y=0$.}
\label{v2}
\end{figure}

%
\begin{figure}
\mbox{
   \epsfysize=3.8in
\epsfbox{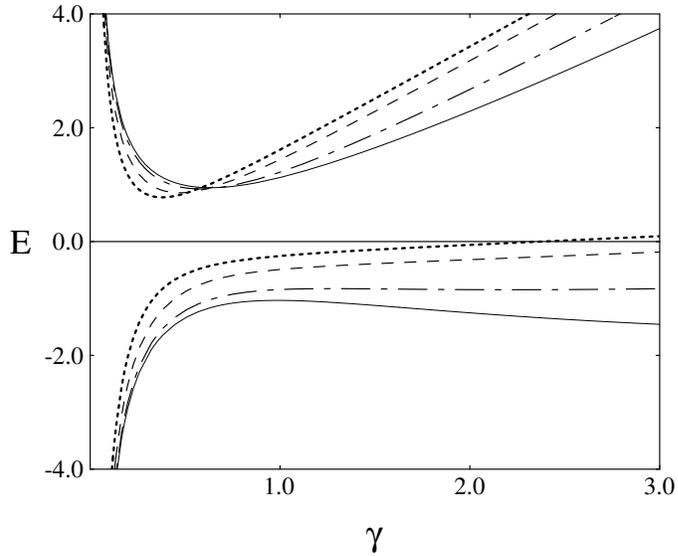}
}
\caption{The 1CS energy $E=E_B$ as a function
of the variational parameter $\gamma$ for different mass ratios $\kappa=m_1/m_2=10$
(solid line), $\kappa=5$ (dot-dashed), $\kappa=2$ (dashed), and $\kappa=1$ (dotted).
In all cases, $y=0$}
\label{v3}
\end{figure}

%
\begin{figure}
\mbox{
   \epsfysize=4.0in
\epsfbox{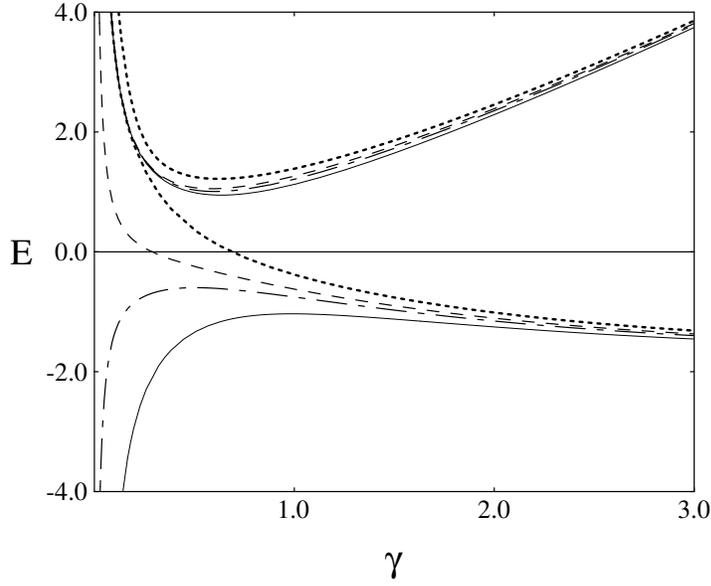}
}
\caption{The 1CS energy $E=E_B$ as a function
of the variational parameter $\gamma$ for different mixing ratios $y=0$ (solid
line), $y=0.4$ (dot-dashed), $y=0.6$ (dashed), and $y=1.0$ (dotted).  In all cases
$m_1/m_2=10$.}
\label{v4}
\end{figure}

%
\begin{figure}
\mbox{
   \epsfysize=4.0in
\epsfbox{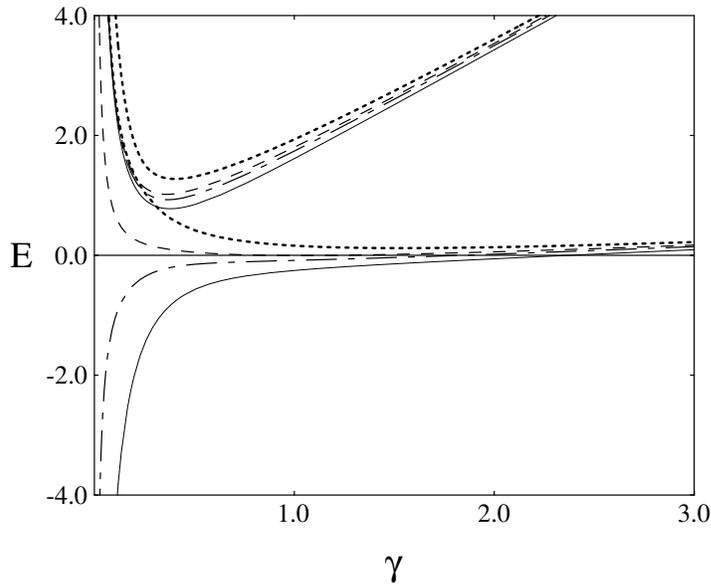}
}
\caption{As in Fig.~11 but with $m_1=m_2$.}
\label{v5}
\end{figure}

Substituting $f$ and $g$ into Eq.~(\ref{der11n}), multiplying the first equation by
$j_0(\gamma r)$ and the second by $j_1(\gamma r)$, and integrating over
$d^3r$ gives the following coupled equations for $f_o$ and $g_o$
\begin{eqnarray}
&&\left(E_B-\left[\sqrt{m_1^2+\gamma^2}-m_1\right]-m_{2}\right)\,f_o +
\gamma\,g_o\nonumber\\
&&\qquad\qquad=\left\{\left(\frac{\sigma c_1}{\gamma}-C(\gamma^2)\right)\left(1-
\tilde{p}^2_1\,(1-2y)\right) - \frac{\sigma c_2}{\gamma}
\tilde{p}^2_1\,(1-2y)\right\}\,{\cal N}\,f_o \equiv S_f\,f_o
\nonumber\\
&&\left( E_B-\left[\sqrt{m_1^2+\gamma^2}-m_1\right]+m_{2}\right)\,g_o
-\gamma\,b\,f_o\nonumber\\ 
&&\qquad\qquad=-\left\{\left(\frac{\sigma c_3}{\gamma}-C(\gamma^2)\right)
\left(1-2y -\tilde{p}_1^2\right)  -\frac{\sigma c_4}{\gamma}
\tilde{p}^2_1\right\}\,{\cal N}\,g_o \equiv -S_g\,g_o \, ,
\label{r7}
\end{eqnarray}
where
\begin{equation}
\begin{array}{ll}
c_1= {\displaystyle\frac{\int_0^\pi x^3dx j_0^2(x)}{\int_0^\pi x^2dx j_0^2(x)} =
1.571} & 
c_3= {\displaystyle\frac{\int_0^{n_1} x^3dx j_1^2(x)}{\int_0^{n_1}
x^2dx j_1^2(x)} = 2.659} \cr
& \cr 
c_2= {\displaystyle\frac{\int_0^\pi x dx j_0^2(x)}{\int_0^\pi
x^2dx j_0^2(x)} = 0.776 }\qquad & 
c_4= {\displaystyle\frac{\int_0^{n_1} x dx j_1^2(x)}
{\int_0^{n_1} x^2 dx j_1^2(x)}=0.412 }
\end{array}
\end{equation}
and
\begin{eqnarray}
b={\displaystyle\frac{\int_0^\pi x^2 dx j_1^2(x)}{\int_0^{n_1} x^2 dx j_1^2(x)} =
0.734}
\,  .
\end{eqnarray}
Solving Eq.~(\ref{r7}) gives an estimate for the eigenvalues $E_B$ as a function of
$\gamma$, related to the size of the state
\begin{eqnarray}
E_B=\sqrt{m_1^2+\gamma^2}-m_1+{1\over2}(S_f-S_g)\pm\sqrt{{1\over4}(2m_2
+S_f+S_g)^2+\gamma^2\,b} 
\,  ,  \label{e8}
\end{eqnarray}
where $S_f$ and $S_g$ were defined in Eq.~(\ref{r7}).  These energy surfaces for a variety of
cases are shown in Figs.~\ref{v1}--\ref{v5}.  In all of these cases we chose
$\sigma=0.2$ GeV$^2$ and $m_2=0.325$ GeV.  We will now discuss some of the
interesting features of these solutions.

\begin{table}
\begin{center}
\begin{minipage}{5.0in}
\caption{Comparison of the exact and estimated solutions for the Dirac and 1CS
equations. All energies are in GeV, and the symbol -- indicates that
there is no stable solution.}
\label{Variational}
\vspace*{0.1in}
\begin{tabular}{cc|c|cc|c|cc}
 \multicolumn{2}{c|}{parameters}& \multicolumn{3}{c|}{positive
energy} &
\multicolumn{3}{c}{negative energy} \\
\tableline
& & exact & \multicolumn{2}{c|}{estimate} &  exact & 
\multicolumn{2}{c}{estimate}  \\
$m_1/m_2$ & {$y\;$} &  $E_1$ & $E$ & $\gamma\;$ &
$E_{-1}$ & $E$ & $\gamma\;$ \\
\tableline 
\multicolumn{8}{c}{Dirac} \\
\tableline 
$\infty$ & 0.0$\;$ & 0.976$\;$ & 0.950  &  0.715$\;$ &-1.249  & -1.226 &
0.859 $\;$ \\  
& 0.4$\;$ & 1.028$\;$ & 1.014	 &  0.673$\;$ & -0.660   & -0.650 &
0.463 $\;$ \\
\tableline 
\multicolumn{8}{c}{One Channel Spectator} \\
\tableline 
10 & 0.0$\;$ & 0.964 & 0.946$\;$ &  0.635$\;$  & -1.091 & -1.034   &
0.988 $\;$ \\
   &  0.4$\;$ & 1.013 &  1.007$\;$ &  0.603$\;$ & -0.619 &   -0.598 &
0.505$\;$\\ 
5 &  0.0$\;$ & 0.940  &  0.926$\;$  & 0.579$\;$  & -0.936  &  -0.828  & 
1.272 $\;$\\
  &  0.4$\;$ & 0.992  & 0.992 $\;$  &   0.552$\;$ & -0.548 &  -0.532
&  0.563 $\;$\\ 
2 &  0.0$\;$ & 0.857  & 0.857 $\;$  &  0.471$\;$  & -0.607 &  --  & 
-- $\;$ \\
  &  0.4$\;$ & 0.928 &  0.952 $\;$  &  0.452 $\;$ & -- &  --  &  --
$\;$\\ 
1 &  0.0$\;$ & 0.745 & 0.777 $\;$ &   0.379 $\;$ & -0.330  & --  & -- $\;$
\\
  &  0.4$\;$ & 0.853 &  0.928 $\;$  &  0.367 $\;$ & -- & --  & -- $\;$
\\
\end{tabular}
\end{minipage}
\end{center}
\end{table}

Note that the solutions (\ref{e8}) are always real, and that as $\gamma\to0$
\begin{eqnarray}
E_B\to \frac{\sigma}{2\gamma}\left(c_1-c_3(1-2y)\pm |c_1+c_3(1-2y)|\right)
\end{eqnarray}
Hence the positive energy solution always approaches $+\infty$ as $\gamma\to0$, but the
negative energy solution goes like
\begin{eqnarray}
E_B^-\to\frac{\sigma}{\gamma}\displaystyle{\cases{-c_3(1-2y) & if
$y<0.795=\displaystyle{\frac{c_1+c_3}{2c_3}}$ \cr 
& \cr
c_3(2y-1) \qquad & if $y>0.795 \, ,$\cr}}
\end{eqnarray}
and becomes positive for $y>1/2$, as shown in
Figs.~\ref{v4} and \ref{v5}.  This is
a sign of instability.  When $y>1/2$ the positive energy states cannot be stable because they
may always reduce their energy by tunneling through to a negative energy surface and sliding
down to $-\infty$.

A similar problem may occur at large $\gamma$, but because our estimates are less reliable
here (we neglected the wave function tails which are more important at
large $\gamma$) we can draw no firm conclusion.  As $\gamma\to\infty$,
\begin{eqnarray}
E_B&&\to \gamma\left(1\pm \sqrt{b}\right) \, ,
\end{eqnarray}
and because $b<1$ the negative energy solutions also become positive at 
large $\gamma$.  This feature sets in at lower values of $\gamma$ as the
mass ratio $m_1/m_2$ decreases, as is shown in Fig.~\ref{v3}.  In fact we do
note that the numerical solutions for the negative energy states are
unstable for small values of $m_1/m_2$, but we see no sign of instability in
the positive energy solutions for small values of $y$ and all values of $m_1/m_2$.

Finally, a comparison between these estimates and exact solutions for the ground state are
summarized in Table~\ref{Variational}.  Note that Eq.~(\ref{der11n}) does a credible job of
explaining the trends, all of which can be understood qualitatively from examination of the
figures.

Before leaving the discussion of the 1CS equation, we comment on two features of our estimates
due to the presence of the ``constant'' term $C(p^2)$ of relativistic origin [recall
Eq.~(\ref{eq8})].  First, note that the positive energy 1CS solutions approach the Dirac limit
as $m_1\to\infty$ from {\it below\/} instead of from above, as would have been suggested by our
analysis of the free particle case.  (Note the comparison in
Fig.~\ref{v2}.)   Even though the energy factor
$\left[E_1-m_1\right]$ is positive, the term $C$ is negative and is just a bit larger,
giving the observed behavior.  Second, the term $-C$ becomes more negative with
decreasing mass ratio, explaining the drop in the binding energy as $m_1/m_2$
decreases to unity.

%
\begin{figure}
\mbox{
   \epsfysize=3.0in
\epsfbox{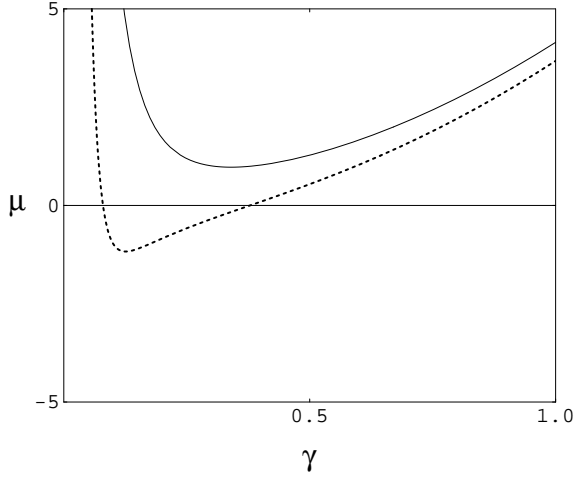}
}
\caption{The square of the bound state mass $\mu^2$ as a function of the
parameter $\gamma$ for the Salpeter equation with a pure scalar
confining interaction ($y=0$).  Solid curve ($m=0.325$), dashed curve
($m=0.1$).}
\label{v6}
\end{figure}

\subsection{Stability of the Salpeter equation}

Applying our technique to the Salpeter equation (\ref{spder2a}) for equal masses 
($m_1=m_2=m$) gives
\begin{eqnarray}
\left(\mu-2E \right)\,f_1=&&w_1\,f_1 + w_2\,f_2 \nonumber\\ 
\left(\mu+2E \right)\,f_2=&&-w_1\,f_2 -w_2\,f_1  
\end{eqnarray}
where
\begin{eqnarray}
&&w_1=\left\{\left(\frac{\sigma
c_1}{\gamma}-C(\gamma^2)\right)\left(\left[1-
\tilde{p}^2\right]^2+4y\,\tilde{p}^2\right) - 2\frac{\sigma c_2}{\gamma}\,
\tilde{p}^2\,(1-2y)\right\}\,{\cal N}\nonumber\\
&&w_2=2\left\{\left(\frac{\sigma c_1}{\gamma}-C(\gamma^2)\right)
\,2\tilde{p}^2\,(1-y) + \frac{\sigma c_2}{\gamma}
\tilde{p}^2\,(1-2y)\right\}\,{\cal N}\, ,
\end{eqnarray}
and we have assumed that $f_1$ and $f_2$ are both S-states.   Hence the estimated
mass is
\begin{eqnarray}
\mu^2= \left(2E+w_1\right)^2-w_2^2\, .
\end{eqnarray}
We have recovered the result that the masses always occur in $\pm$ pairs,
and we see that they may be imaginary if $|w_2|>|2E+w_1|$.

\begin{table}
\begin{center}
\begin{minipage}{3.5in}
\caption{Comparison of the exact and estimated solutions for the Salpeter
equation.  All energies are in GeV and the symbol -- indicates that
there is no stable solution.}
\label{Variationalsal}
\vspace*{0.1in}
\begin{tabular}{cc|c|cc}
\multicolumn{2}{c|}{parameters}  & exact &
\multicolumn{2}{c}{estimate}
\\
$m$  &  $y$ & $E_1^2$ & $E^2$ & $\gamma$ \\  
\tableline 
0.325 & 0.0$\;$ & --    $\;$    & 0.973  $\;$ & 0.340 $\;$  \\
      & 0.4$\;$ & 1.339 $\;$    & 1.537  $\;$ & 0.349 $\;$  \\
      & 0.6$\;$ & 1.510 $\;$    & 1.819  $\;$ & 0.353 $\;$  \\
      & 1.0$\;$ & 1.837 $\;$    & 2.380  $\;$ & 0.361 $\;$  \\
0.650 & 0.0$\;$ & 3.112 $\;$    & 3.217  $\;$ & 0.466 $\;$  \\
0.900 & 0.0$\;$ & 5.235 $\;$    & 5.396  $\;$ & 0.529 $\;$  \\
\end{tabular}
\end{minipage}
\end{center}
\end{table}

First note that as $\gamma\to0$,
\begin{eqnarray}
\mu^2= \left(\frac{\sigma c_1}{\gamma}\right)^2 \, ,
\end{eqnarray}
and as  $\gamma\to\infty$,
\begin{eqnarray}
\mu^2= (2\gamma)^2 \, ,
\end{eqnarray}
so that $\mu^2$ is always large and positive at the extreme values of
$\gamma$, and must have a minimum for some $\gamma$. If this minimum
is {\it negative\/}, the masses will be imaginary (i.e., the state will
be unstable).  This can occur only if $m$ and $y$ are small enough to
satisfy the condition   
\begin{eqnarray}
&&2E +\left\{\left(\frac{\sigma
c_1}{\gamma}-C(\gamma^2)\right)\left(1-6\tilde{p}^2 + \tilde{p}^4 
+8y\,\tilde{p}^2\right) - 4\frac{\sigma c_2}{\gamma}\,
\tilde{p}^2\,(1-2y)\right\}\,{\cal N} \nonumber\\
&&\qquad=2E+w_1-w_2<0 \, . \label{condition}
\end{eqnarray}
If $m=0$ this condition reduces to
to 
\begin{eqnarray}
2\gamma -\frac{4\sigma}{\gamma}\left(c_1 +c_2 -\frac{2}{\pi}\right)
(1-2y)  <0\, .
\end{eqnarray}
Hence the Salpeter equation for $m=0$ is unstable only if $y<1/2$! 
As $m$ increases, this critical value of $y$ decreases.  If $y=0$, our
estimate Eq.~(\ref{condition}) leads to the conclusion that the scalar
Salpeter equation is unstable only if $m<0.18$; for larger values of
$m$ the equation has real roots for all $y$.  This behavior is
illustrated in Fig.~\ref{v6}, which shows that the scalar Salpeter
equation is stable for $m=0.325$ (our standard choice for the quark
mass) and unstable for $m=0.1$.

Unfortunately, our crude estimate Eq.~(\ref{condition}) does not reproduce
the quantitative features of the exact Salpeter solutions as well as it did for
the previous cases.  The comparison between exact and estimated
solutions is given for a few cases in Table \ref{Variationalsal}. 
Note that the qualitative agreement is good, but that we are unable to
``predict'' the the critical mass at which the Salpeter equation
becomes unstable.  The exact solutions tell us that this mass is
around 0.85 GeV, much higher than the estimated value of 0.18.

\section{numerical results}
Now we turn our attention to the numerical solutions for the Dirac, 1CS, and
Salpeter equations.  

Numerical results are obtained by expanding the solutions in
terms of splines, as described in Appendix A.  In this way the integral
equations in momentum space are turned into matrix equations and the problem
reduced to a generalized matrix eigenvalue problem.  Numerical values of the
eigenvectors (expansion coefficients) and the eigenvalues (bound state masses or
binding energies) are obtained, and the wave functions are constructed from the
spline expansion.

\subsection{The Dirac equation}

The Dirac equation is reduced to the system given in Eq.~(\ref{iddr31})
and Eq.~(\ref{iddr31a}) and can be solved numerically on a PC in a 
reasonable length of time.  The antiquark mass was set to $m=0.325$
GeV and the confinement strength $\sigma=0.2$ GeV$^2$.  We looked at
four different values of the vector strength: $y=0.0$
(pure scalar), 0.4, 0.6, and 1.0 (pure vector). The first four
positive and negative energy levels for $y$ values of 0.0, 0.4 and 0.6
are listed in Table~IV for spline ranks of 12, 16, and 20.  The pure
vector case ($y=1.0$) was found to be fully unstable, as predicted by
Fig.~\ref{v1}, and is not listed in the Table.  The eigenvalues, which
for the Dirac equation are the binding energies, are all real and
therefore pass the first stability condition (as described in
Sec.~ID). 

\begin{table}
\begin{center}
\begin{minipage}{6.0in}
\caption{First four positive and negative Dirac energy levels 
for $y$=0.0, 0.4, and 0.6 with spline ranks of 20, 16, and
12.  The energies are in GeV.  The bold face numbers
are unstable states with energies {\it greater\/} than the
stable ground state, as discussed in the text.}\label{TDirac}
\vspace*{0.1in}
\begin{tabular}{r|rrr|rrr|rrr}
& \multicolumn{3}{c|}{$y=0.0$} &
\multicolumn{3}{c|}{$y=0.4$} & \multicolumn{3}{c}{$y=0.6$} \\
 \cline {2-10}  & & & & & & & & &\\[-0.15in] 
Level & SN=20 & SN=16 &  SN=12$\;$ & SN=20 & SN=16 &  SN=12$\;$ & SN=20 &
SN=16 &  SN12 \\[0.15cm]
\tableline
4 $\;$ & 1.945 & 1.945 & 1.946$\;$ & 2.035 & 2.035 & 2.035$\;$ & 2.092 & 2.092  
& 2.093\\
3 $\;$ & 1.695 & 1.695 & 1.695$\;$ & 1.772 & 1.772 & 1.772$\;$ & 1.821 & 1.821 
& 1.821\\
2 $\;$ & 1.394 & 1.393 & 1.393$\;$ & 1.456 & 1.455 & 1.455$\;$ & 1.496 & 1.496
& 1.496\\
1 $\;$ & 0.976 & 0.976 & 0.976$\;$ & 1.028 & 1.028 & 1.028$\;$ & 1.065 & 1.065
& 1.065\\
-1 $\;$ & -1.249 & -1.249 & -1.248$\;$ & -0.660 & -0.660 & -0.660$\;$ & {\bf 2.028} 
& {\bf 1.576} & {\bf 1.120}\\
-2 $\;$ & -1.575 & -1.575 & -1.574$\;$ & -0.781 & -0.781 & -0.780$\;$ & {\bf 1.190} & 0.861
& 0.525\\
-3 $\;$ & -1.839 & -1.839 & -1.838$\;$ & -0.879 & -0.878 & -0.879$\;$ & 0.899 & 0.590
& 0.278\\
-4 $\;$ & -2.067 & -2.067 & -2.078$\;$ & -0.963 & -0.963 & -0.964$\;$ & 0.692 & 0.396
& 0.090\\
\end{tabular}
\end{minipage}
\end{center}
\end{table}

Of the four cases studied, only the negative energy levels for
the $y>1/2$ cases (i.e., $y$=0.6 and 1.0) vary significantly with the
spline rank, as shown in Table~IV.  This violates the second of the
stability conditions defined in Sec.~ID.  Furthermore, the bold face
values in Table~IV highlight unstable states whose eigenvalues are
greater than the positive ground state, and hence the $y>1/2$
equations also violate the third stability condition.  These
unstable states were identified and tracked with changing spline
number by looking at their momentum space structure, as discussed
below.   
%
\begin{figure}
\begin{center}
\mbox{
   \epsfysize=3.0in
\epsfbox{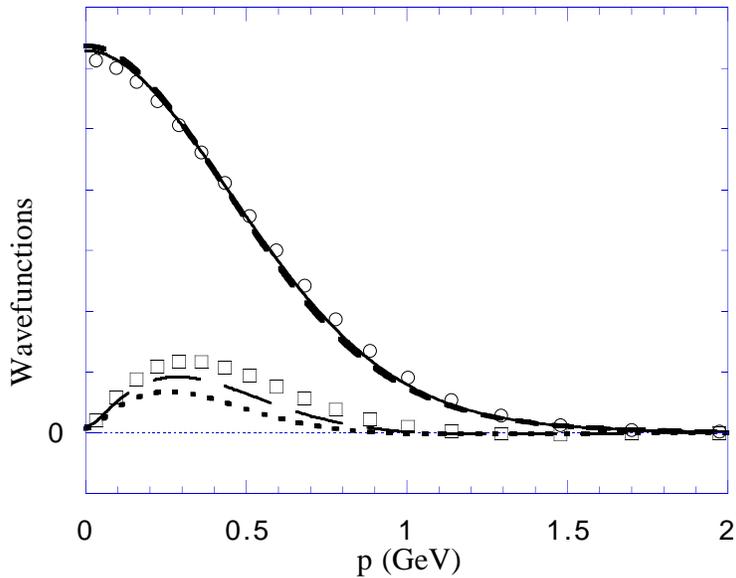}
}
\end{center}
\caption{Dirac positive ground state solutions for three values of the 
vector strength $y$:  $y=0.0$,
$E_{1}$=0.976 GeV (circles and squares);  $y=0.4$, $E_{1}$=1.028 GeV (solid and long
dashed lines); and for $y=0.6$, $E_{1}$=1.065 GeV (heavy short dashed and dotted
lines).}
\label{fig6a}
\end{figure}
%

%
\begin{figure}
\vspace{0.5in}
\begin{center}
\mbox{
   \epsfysize=3.0in
\epsfbox{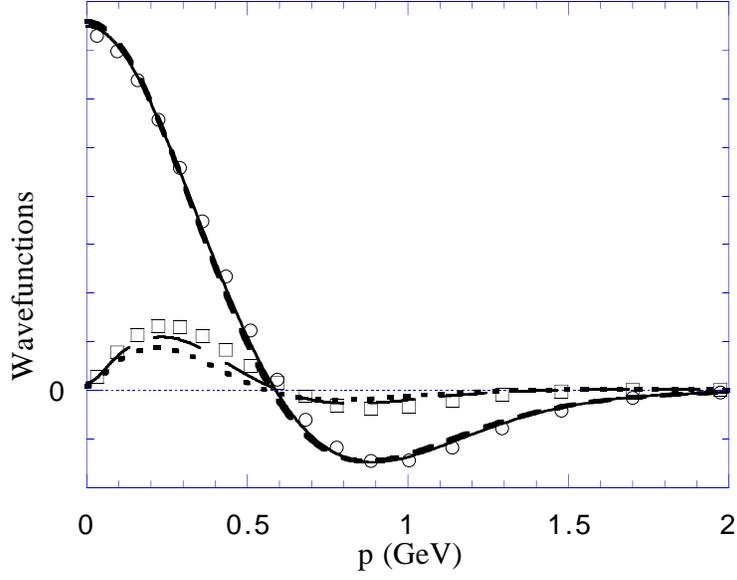}
}
\end{center}
\caption{Dirac positive first excited state solutions for $y=0.0$, $E_{2}=1.394$ GeV
(circles and squares), for $y=0.4$, $E_{2}$=1.456 GeV (solid and long dashed lines), and
for $y=0.6$, $E_{2}$=1.496 GeV (heavy short dashed and dotted lines).}
\label{fig6b}
\end{figure}
%
       
%
\begin{figure}
\begin{center}
\mbox{
   \epsfysize=3.0in
\epsfbox{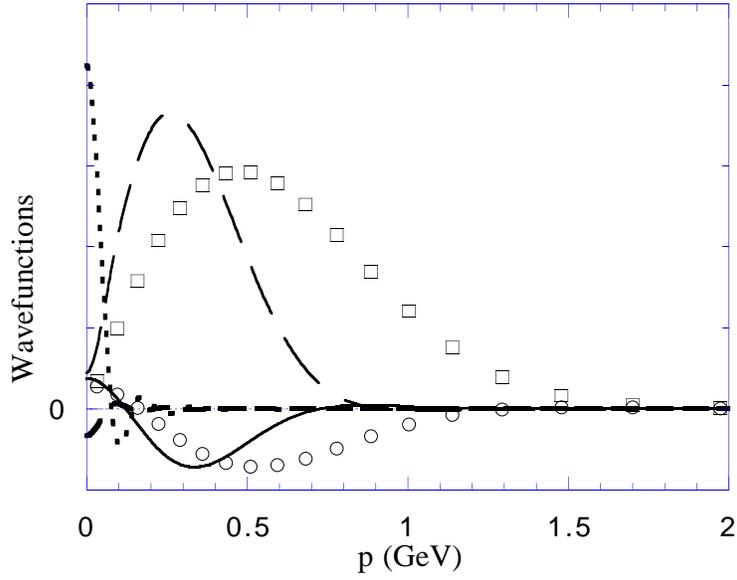}
}
\end{center}
\caption{Dirac negative ground state solutions for $y=0.0$, $E_{-1}=-1.249$ GeV 
(circles and squares), for $y=0.4$, $E_{-1}$=-0.660 GeV (solid
and long dashed lines), and for $y=0.6$, $E_{-1}=2.028$ GeV (heavy short dasheded and
dotted lines).}
\label{fig6c}
\end{figure}

The Dirac wave functions are shown in Figs.~\ref{fig6a}--\ref{fig6c}. 
Fig.~\ref{fig6a} gives the positive energy ground states,
Fig.~\ref{fig6b} the first positive energy excited states, and
Fig.~\ref{fig6c} the negative energy ground states.  By comparing the
solutions for the states with $y<1/2$ (which are known to be stable)
with the $y=0.6$ solutions, we conclude that (i) the positive energy
$y=0.6$ state has a structure identical to the other positive energy
states, and hence appears to be stable (as already suggested by the
stability of the eigenvalue shown in Table
\ref{TDirac}), but (ii) the negative energy $y=0.6$ ground state, shown
in Fig.~\ref{fig6c}, has a radically different structure (similar to a
momentum space delta function) showing that it is indeed unstable. 
All of the $y=1.0$ solutions (not shown in the figures) have a
behavior similar to the negative energy $y=0.6$ solution,
confirming that they are unstable.       

The apparant stability of the $y=0.6$ positive energy solution
differs from expectations based on the discussion in Sec.~II and 
examination of Fig.~\ref{v1}.  We expect all positive energy
solutions for $y>1/2$ to be unstable, but as Fig.~\ref{v1} shows, the
positive and negative energy surfaces actually {\it overlap\/} in the
$y=1.0$ case but remain {\it clearly separated\/} for the $y=0.6$
case.  This suggests that the instability of the $y=0.6$ positive
energy state is hard to observe numerically because the distance
between the positive and negative energy surfaces is large and the
``leakage'' from positive to negative energy is very small (also
suggested by Fig.~\ref{figd}).  Presumably a more precise numerical
calculation would uncover some instability in the positive energy
$y=0.6$ case, but this further calculation is not needed because the
overlap of the positive and negative energy spectrum (condition 3) is
already a sign of the instability.  

We conclude that the fourth stability condition largely reinforces the
conclusions we have already drawn, but that it should be used in conjunction
with the other three.  The stability of a {\it single\/} state cannot easily be
determined solely by tracking (with changing spline number) its behavior.  A reliable
conclusion requires the examination of the entire spectrum, with particular
attention to condition 3.

\subsection{The one-channel spectator equation}
As in the Dirac case the antiquark mass will be set to $m_{2}=0.325$
GeV and the confinement strength to $\sigma=0.2$ GeV$^2$.  We will
present results for heavy quark masses $m_1=\kappa\,m_2$ with the mass
ratio $\kappa=10$, 5, 2, and 1.  In order to compare the 1CS results
to those obtained from the Dirac equation, we define an effective
Dirac-like binding energy $E_D$ using the relation    
\begin{equation}
\mu = E_{D}+m_{1}\, 
\label{mu1}
\end{equation}
where $\mu$ is the mass eigenvalue obtained from the 1CS equation.
This relation insures that the effective 1CS binding energy must
approach the Dirac binding energy as $m_{1}\to\infty$.  Tables~V and
VI give these  effective  binding energies (instead of the bound
state masses).

\begin{table}
\begin{center}
\begin{minipage}{6.0in}
\caption{First four positive and negative energy levels for the 1CS
equation for mass ratios $\kappa$=5.0 and 10.0 and vector strength
$y$=0.0 and 0.4.  Here
$E_D$ is shown in GeV and solutions for spline ranks of 20 and
12 are compared.}\label{T1CSf5e1}
\begin{tabular}{r|rr|rr|rr|rr}
& \multicolumn{2}{c|}{$y=0.0\;$ $\kappa=5.0$ } &
\multicolumn{2}{c|}{$y=0.0\;$ $\kappa=10.0$} &
\multicolumn{2}{c|}{$y=0.4\;$ $\kappa=5.0$} &
\multicolumn{2}{c}{$y=0.4\;$
$\kappa=10.0$}\\
 \cline {2-9}  & & & & & & & & \\[-0.15in] 
Level & SN=20 & SN=12$\;$ & SN=20 & SN=12$\;$ & SN=20 & SN=12$\;$ & SN=20 & SN12
\\[0.15cm]
\tableline
4 $\;$ & 2.109 & 2.113$\;$ & 2.073 & 2.078$\;$ & 2.225 & 2.227$\;$ & 2.165 & 2.168
\\
3 $\;$ & 1.808 & 1.808$\;$ & 1.783 & 1.783$\;$ & 1.898 & 1.899$\;$ & 1.858 & 1.858
\\
2 $\;$ & 1.443 & 1.443$\;$ & 1.435 & 1.435$\;$ & 1.509 & 1.509$\;$ & 1.495 & 1.494
\\
1 $\;$ & 0.940 & 0.939$\;$ & 0.964 & 0.964$\;$ & 0.992 & 0.992$\;$ & 1.013 & 1.013
\\
-1 $\;$ & -0.936 & -0.936$\;$ & -1.091 & -1.090$\;$ & -0.548 & -0.569$\;$ & -0.619 &
-0.619 \\
-2 $\;$ & -1.084 & -1.084$\;$ & -1.333 & -1.332$\;$ & -0.570 & -0.607$\;$ & -0.715 &
-0.715 \\
-3 $\;$ & -1.173 & -1.170$\;$ & -1.511 & -1.515$\;$ & -0.600 & -0.637$\;$ & -0.786 &
-0.785 \\
-4 $\;$ & -1.233 & -1.259$\;$ & -1.650 & -1.642$\;$ & -0.630 & -0.675$\;$ & -0.841 &
-0.848 \\
\end{tabular}
\end{minipage}
\end{center}
\end{table}

Note that results for the equal mass ($\kappa=1.0$) 1CS equation are
included only for comparison because the 1CS equation should {\it not\/} be
used for equal mass systems.  If the equal mass particles are identical (as in
$NN$ scattering) the equation must be symmetrized in order to
preserve the Pauli principle.  Even if the equal mass particles are not identical,
as for the $q\bar{q}$ pairs discussed in this paper, the equation must still be
symmetrized to insure charge conjugation invariance.  Furthermore, for bound
states with a very small mass (e.g., the pion) the symmetrized {\it two
channel\/} spectator equation defined in Ref.~\cite{gross} should be used.

\begin{table}
\begin{center}
\begin{minipage}{6.0in}
\caption{First four positive and negative energy levels for the 1CS
equation for the mass ratio $\kappa$=1.0 and vector strength
$y$=0.0 and 0.4.  Here
$E_D$ is shown in GeV and solutions for spline ranks of 24, 20, 16,
and 12 are compared.}\label{T1CSf1}
\begin{tabular}{r|rrrr|rrrr}
& \multicolumn{4}{c|}{$y=0.0$} & \multicolumn{4}{c}{$y=0.4$} \\
 \cline {2-9}  & & & & & & & & \\[-0.15in] 
Level & SN=24 & SN=20 & SN=16 &  SN=12$\;$ & SN=24 & SN=20 & SN=16 &  SN12 \\[0.15cm]
\tableline
4 $\;$ & 1.881 & 1.881 & 1.881 & 1.881$\;$ & 2.222 & 2.222 & 2.222 & 2.223
\\
3 $\;$ & 1.630 & 1.630 & 1.630 & 1.632$\;$ & 1.884 & 1.884 & 1.884 & 1.884
\\
2 $\;$ & 1.294 & 1.293 & 1.293 & 1.293$\;$ & 1.461 & 1.461 & 1.461 & 1.461
\\
1 $\;$ & 0.745 & 0.745 & 0.745 & 0.745$\;$ & 0.853 & 0.853 & 0.853 & 0.853
\\
-1 $\;$ & -0.329 & -0.330 & -0.331 & -0.334$\;$ & {\bf 0.933} & 0.724 & 0.508 & 0.284
\\
-2 $\;$ & -0.331 & -0.332 & -0.335 & -0.341$\;$ & 0.727 & 0.527 & 0.326 & 0.122
\\
-3 $\;$ & -0.334 & -0.337 & -0.342 & -0.354$\;$ & 0.577 & 0.387 & 0.196 & 0.005
\\
-4 $\;$ & -0.338 & -0.343 & -0.353 & -0.379$\;$ & 0.454 & 0.272 & 0.091 & -0.087
\\
\end{tabular}
\end{minipage}
\end{center}
\end{table}

The eigenvalues are real for all values of the vector strength $y$ and
the mass ratio $\kappa$ (condition 1).  However only systems with a
vector strength less than 1/2  (0.0 and 0.4) have stable eigenvalues
(condition 2).  Cases which fail the first two
stability conditions ($y=0.6$ and 1.0) are not listed in the
eigenvalue tables.  Table \ref{T1CSf5e1} shows the
eigenvalues for mass ratios $\kappa$=5.0 and 10.0.  These cases
are very similar to the Dirac cases, and the table shows that
in all cases the spectra satisfy condition 3 (no overlap of the
positive and negative energy sectors).  Table \ref{T1CSf1} shows the
eigenvalues for the equal mass case ($\kappa$=1.0).  Note that condition
3 is violated for $y$=0.4; at a spline rank of 24 the negative
energy state (shown in bold face) crosses into the positive energy
sector.  In the equal mass case only the pure scalar interaction is
stable.  The binding energies for $\kappa=2.0$ (not shown
in the tables) exhibit the same behavior as for $\kappa=1.0$.

%
\begin{figure}
\begin{center}
\mbox{
   \epsfysize=3.0in
\epsfbox{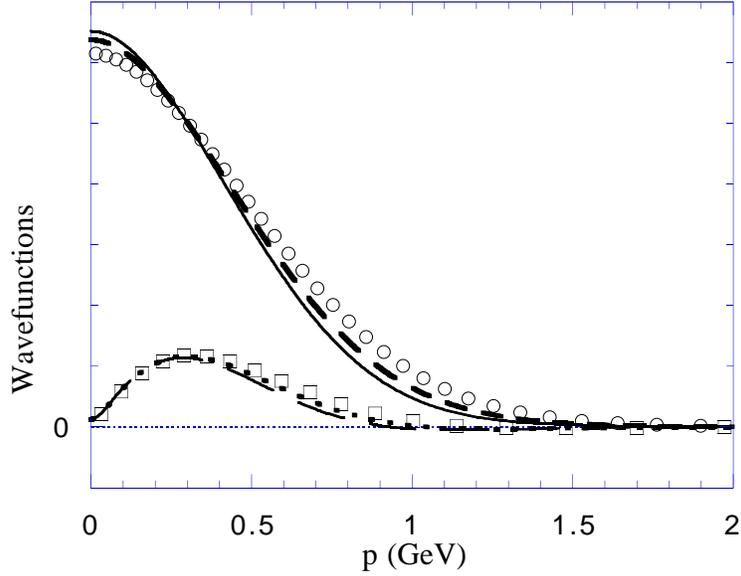}
}
\end{center}
\caption{Positive ground state soutions for the quasirelativistic 1CS equation with a
pure scalar interaction. The solid and long dashed lines are for $\kappa$=5.0,
$E_{1}$=0.940 GeV; the heavy short dashed and dotted lines are for $\kappa$=10.0,
$E_{1}$=0.964 GeV.  The scalar ground state Dirac solution for $E_{1}=0.976$ GeV is shown
for comparison (circles and squares).}
\label{fig17a}
\end{figure}

%
\begin{figure}
\begin{center}
\mbox{
   \epsfysize=3.0in
\epsfbox{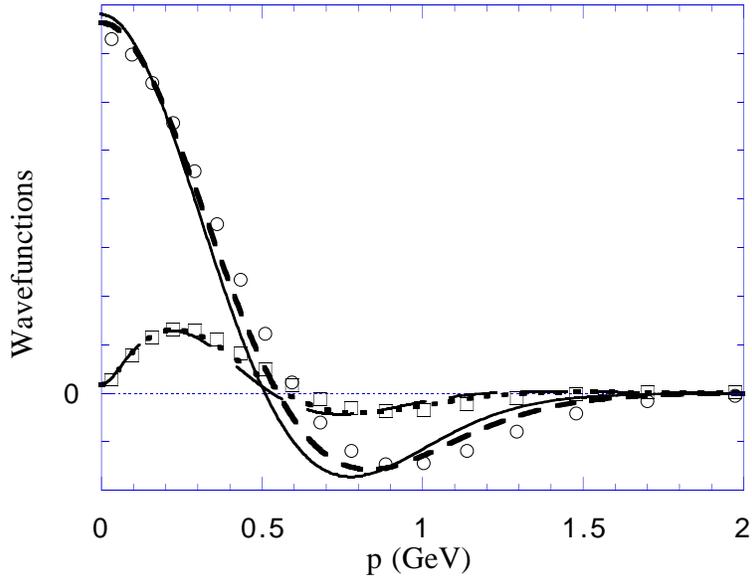}
}
\end{center}
\caption{Positive first excited state solutions labeled as in Fig~17.  Here
the $\kappa=5.0$ solution has an energy of $E_{2}=1.443$ GeV and the $\kappa=10.0$
solution an energy of $E_{2}=1.435$ GeV compared to the Dirac energy of $E_{2}=1.394$
GeV.}
\label{fig18}
\end{figure}

%
\begin{figure}
\begin{center}
\mbox{
   \epsfysize=3.0in
\epsfbox{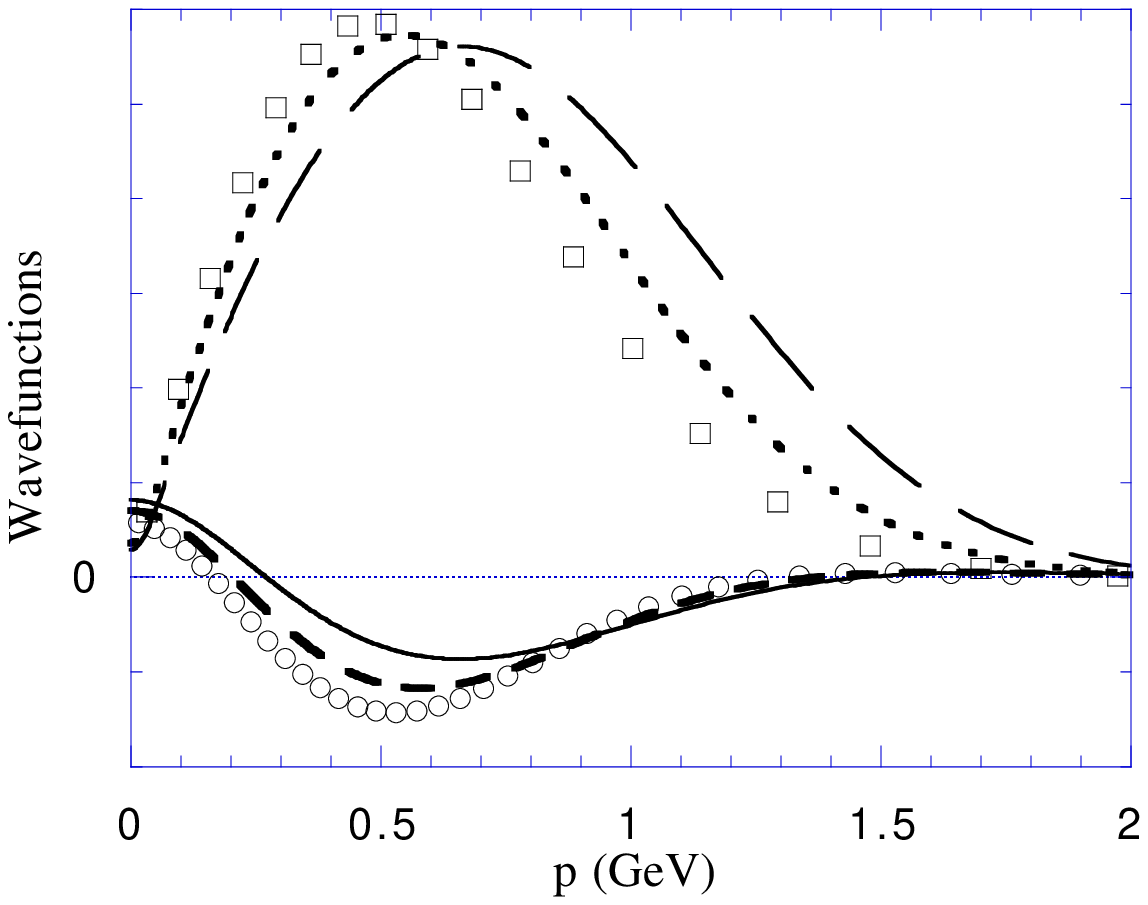}
}
\end{center}
\caption{Negative ground state soutions labeled as in Fig~17.  Here
the $\kappa=5.0$ solution has an energy of $E_{-1}=-0.936$ GeV and the
$\kappa=10.0$ solution an energy of $E_{-1}=-1.091$ GeV compared to the Dirac energy
of $E_{-1}=-1.249$ GeV.}
\label{fig19}
\end{figure}
%

%
\begin{figure}
\begin{center}
\mbox{
   \epsfysize=3.0in
\epsfbox{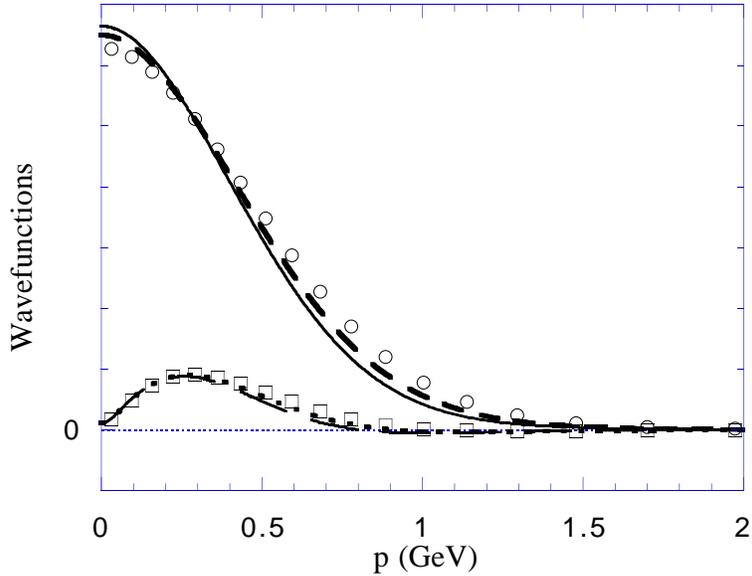}
}
\end{center}
\caption{Positive ground state solutions of the quasirelativistic 1CS equation
with a mixed scalar and vector interaction ($y=0.4$) for two mass ratios
$\kappa$.  The solid and long dashed lines are for $\kappa=5.0$,
$E_{1}=0.992$ GeV, and the heavy short dashed and dotted lines are for $\kappa=10.0$,
$E_{1}=1.013$ GeV. The circles and squares show the solution for the Dirac equation with
$E_1=1.028$ GeV. }
\label{fig23}
\end{figure}
%
%
\begin{figure}
\begin{center}
\mbox{
   \epsfysize=3.0in
\epsfbox{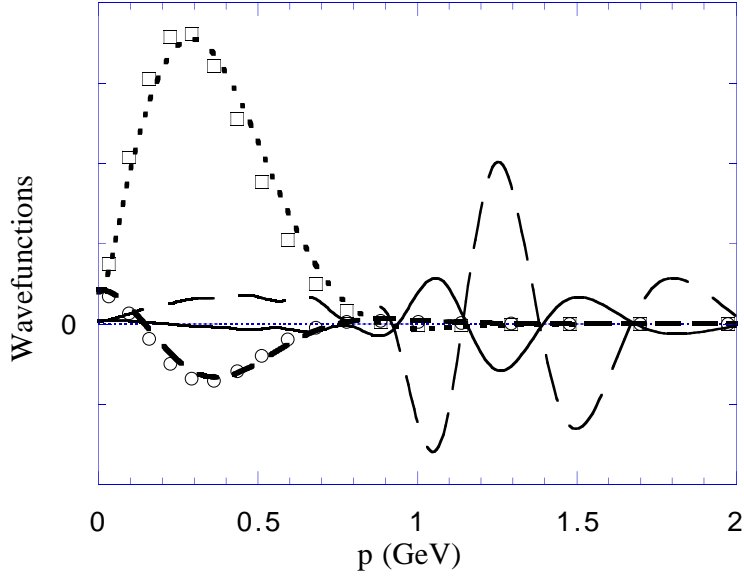}
}
\end{center}
\caption{Negative ground state solutions of the quasirelativistic 1CS equation for
$y=0.4$ labeled as in previous figure.  Here $\kappa=5.0$,
$E_{-1}=-0.548$ GeV and $\kappa=10.0$, 
$E_{-1}=-0.619$ GeV.  The comparison Dirac level has energy $E_{-1}=0.660$ GeV. }
\label{fig24}
\end{figure}
%
%
\begin{figure}
\begin{center}
\mbox{
   \epsfysize=3.0in
\epsfbox{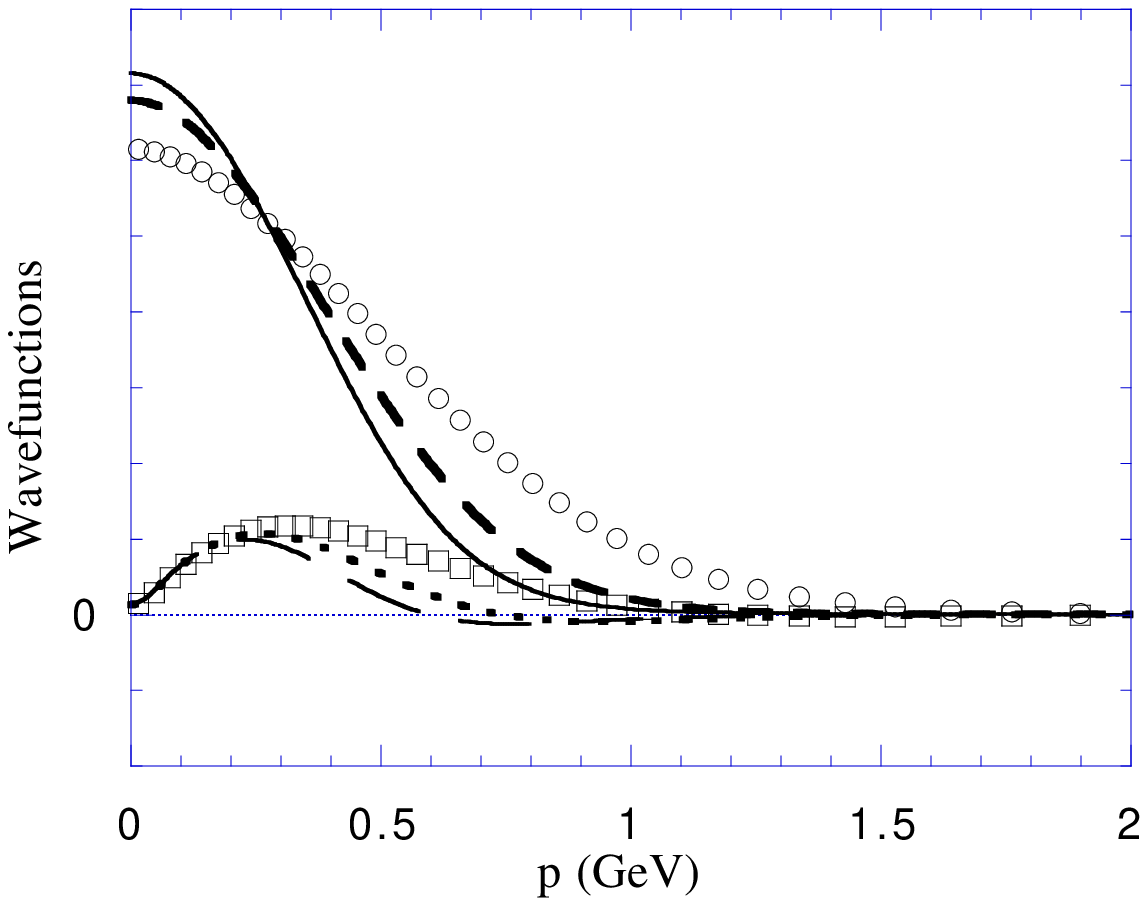}
}
\end{center}
\caption{Positive ground state soutions for the quasirelativistic 1CS equation with a
pure scalar interaction. The solid and long dashed lines are for $\kappa$=1.0,
$E_{1}$=0.745 GeV; the heavy short dashed and dotted lines are for $\kappa$=2.0,
$E_{1}$=0.857 GeV.  The scalar ground state Dirac solution for $E_{1}=0.976$ GeV is shown
for comparison (circles and squares).}
\label{fig17}
\end{figure}
%
%
\begin{figure}
\begin{center}
\mbox{
   \epsfysize=3.0in
\epsfbox{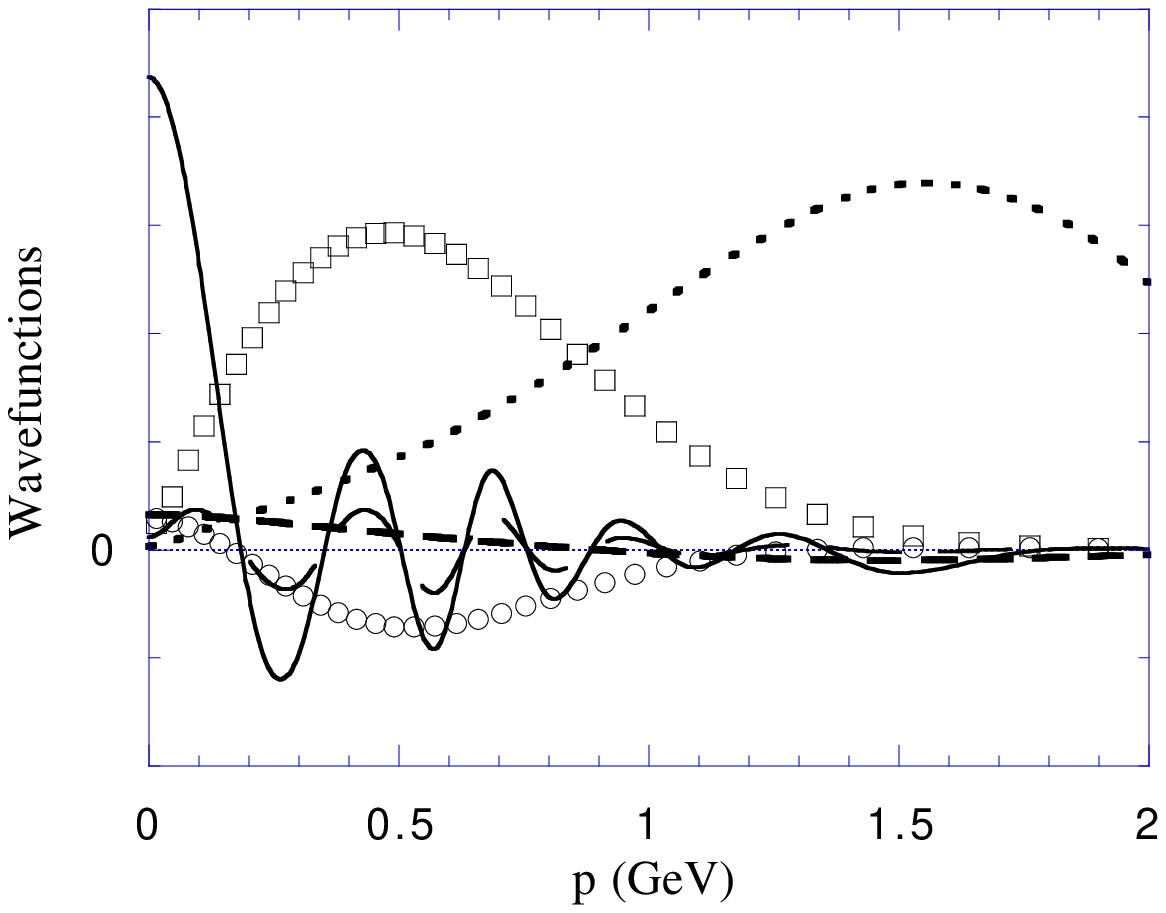}
}
\end{center}
\caption{Negative ground state solutions labeled as in previous figure.  Here
the $\kappa=1.0$ solution has an energy of $E_{-1}=-0.330$ GeV and the
$\kappa=2.0$ solution an energy of $E_{-1}=-0.607$ GeV compared to the Dirac 
energy of $E_{-1}=-1.249$ GeV.}
\label{fig30}
\end{figure}

Wave functions for the 1CS equation are shown in
Figs.~\ref{fig17a}--\ref{fig30}.  In Figs.~\ref{fig17a}--\ref{fig19} the
wave functions for a large mass ratio and a pure scalar confinement are 
compared with the Dirac solutions.  Both the positive and negative states
for these systems are completely stable and very similar to the corresponding
Dirac solutions.  We also observe how the 1CS binding energies approach the
Dirac values as $\kappa$ is increased. 

Figs.~\ref{fig23} and \ref{fig24} show the wave functions for large mass
ratios and a vector strength of 0.4. For $\kappa$=10.0 the system is  once
again totally stable, while for $\kappa$=5.0 only the positive states are
stable. In this case the instability of the negative energy states is {\it
not\/} accompanied by violation of condition 3; the only indication of instability is
the variation of the negative energy levels with spline rank (condition 2), as shown
in Table \ref{T1CSf5e1}.  In this case the structure (condition 4) reinforces
condition 2, and we have a first example of a system where the
positive energy solutions are stable and the negative energy ones are not.

The positive and negative ground states for $\kappa=1.0$ and 2.0 are shown in
Figs.~\ref{fig17} and \ref{fig30}.  Note that the positive energy states are
stable while the negative energy ones are not.  Here the instability
of the negative energy states is only apparant from an examination of the
structure of the wave functions; neither condition 2 (variation of the energy
with spline rank) nor condition 3 (penetration of the positive energy
sector) seems to occur.

In conclusion, the 1CS system {\it becomes more stable\/} as the vector
strength is decreased and the mass of the heavy quark is increased.  This will
be summarized further at the end of this section.

\subsection{The Salpeter Equation}

\begin{figure}
\begin{center}
\mbox{
   \epsfysize=3.0in
\epsfbox{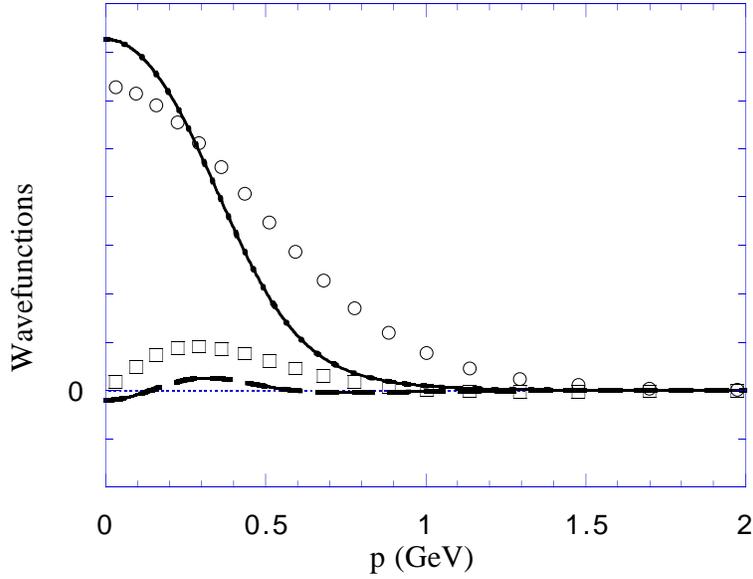}
}
\end{center}
\caption{Positive and negative ground state solutions for the $y=0.4$ quasirelativistic
equal mass Salpeter equation, $\mu_{1}$=1.157 GeV (solid and long dashed lines) 
and $\mu_{-1}$=-1.157 GeV (heavy short dashed and dotted lines).  The positive ground state
Dirac solutions for $y$=0.4, $E_{1}$=1.028 GeV (circles and squares) are shown for
comparison.}
\label{fig25}
\end{figure}

\begin{figure}
\begin{center}
\mbox{
   \epsfysize=3.0in
\epsfbox{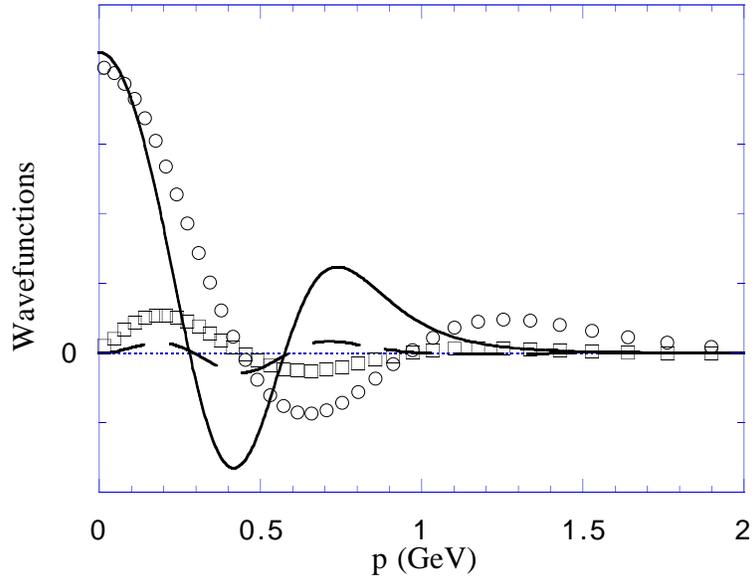}
}
\end{center}
\caption{The positive second excited state solutions for the $y$=0.4 
equal mass Salpeter equation, $\mu_{3}=2.094$ GeV (solid and long dashed lines) are
compared to the second positive excited state  Dirac solution for $y=0.4$,
$E_{3}$=1.772 GeV (circles and squares).}
\label{fig26}
\end{figure}

\begin{figure}
\begin{center}
\mbox{
   \epsfysize=3.0in
\epsfbox{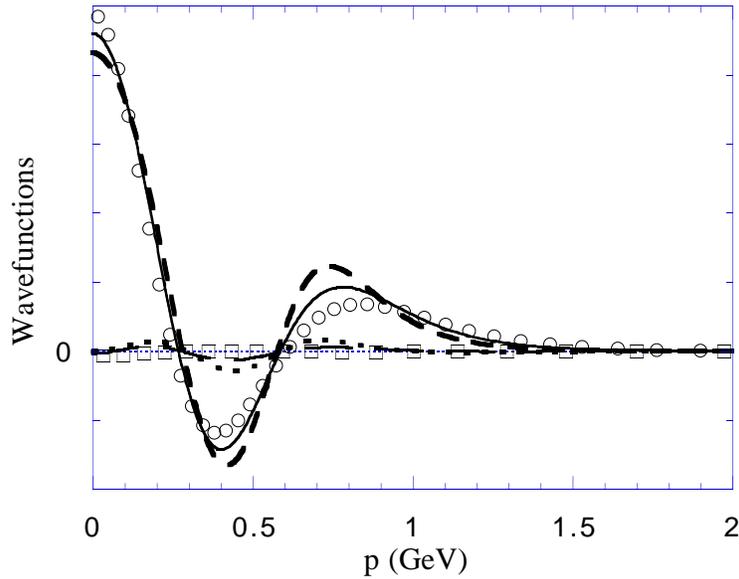}
}
\end{center}
\caption{Positive second excited state solutions for the  Salpeter
equation for a variety of scalar/vector mixings:  pure vector $y$=1.0, 
$\mu_{3}$=2.565 GeV (circles and squares);
$y$=0.6, $\mu_{3}$=2.284 GeV (solid and long dashed lines); and $y$=0.4 $\mu_{3}$=2.094
GeV (heavy short dashed and dotted lines). }
\label{fig27}
\end{figure}

The use of pure scalar confinement with the Salpeter equation gives the first
example of instability due to the mass eigenvalues becoming complex
(condition 1).  Actually, the eigenvalues become pure imaginary because the
mass squared is real and negative.    This situation is accompanied
by a very rapid variation of $\mu^2$ with spline rank, as shown in Table
\ref{TSalpeter}.  However, for $y$=0.4 the tabulated spectra do not vary with
the spline rank, and these states are stable, as shown in
Figs.~\ref{fig25}--\ref{fig27}.  Fig.~\ref{fig25} also shows that the
wave functions for positive and negative energies are identical provided
$\psi_{1a}\leftrightarrow\psi_{2a}$.  This is a further consequence of the 
symmetry of the Salpeter equation which produces pairs of eigenvalues
with the same magnitude and opposite signs.    

\begin{table}
\begin{center}
\begin{minipage}{6.0in}
\caption{Square of the mass ($\mu^2$ in GeV$^2$) of the first four levels
of the Salpeter equation for $y$=0.0 and 0.4 with various spline
ranks.}\label{TSalpeter}
\begin{tabular}{r|rr|rrr}
& \multicolumn{2}{c|}{$y=0.0$} &
\multicolumn{3}{c}{$y=0.4$} \\
 \cline {2-6}  & & & & & \\[-0.15in] 
Level & SN=20 &  SN=12$\;$ & SN=20 & SN=16 &  SN=12 \\[0.15cm]
\tableline
4 $\;$ & 0.685 & 2.173$\;$ & 5.632 & 5.632 & 5.674
\\
3 $\;$ & -1.074 & 1.538$\;$ & 4.383 & 4.383 & 4.385
\\
2 $\;$ & -3.869 & 0.931$\;$ & 2.977 & 2.977 & 2.976
\\
1 $\;$ & -8.705 & -0.051$\;$ & 1.339 & 1.339 & 1.339
\\
\end{tabular}
\end{minipage}
\end{center}
\end{table}

The two figures, Fig.~\ref{fig25} (ground state) and Fig.~\ref{fig26} (second excited
state), demonstrate that these Salpeter systems have solutions comparable to
their Dirac counterparts.  In addition, Fig.~\ref{fig27} illustrates that the 
$y$=0.6 and 1.0 solutions are indeed stable by showing that they have the
correct structure with the right number of nodes for a second excited
state.   

%
%
%
%

%
\begin{figure}
\begin{center}
\mbox{
   \epsfysize=3.0in
\epsfbox{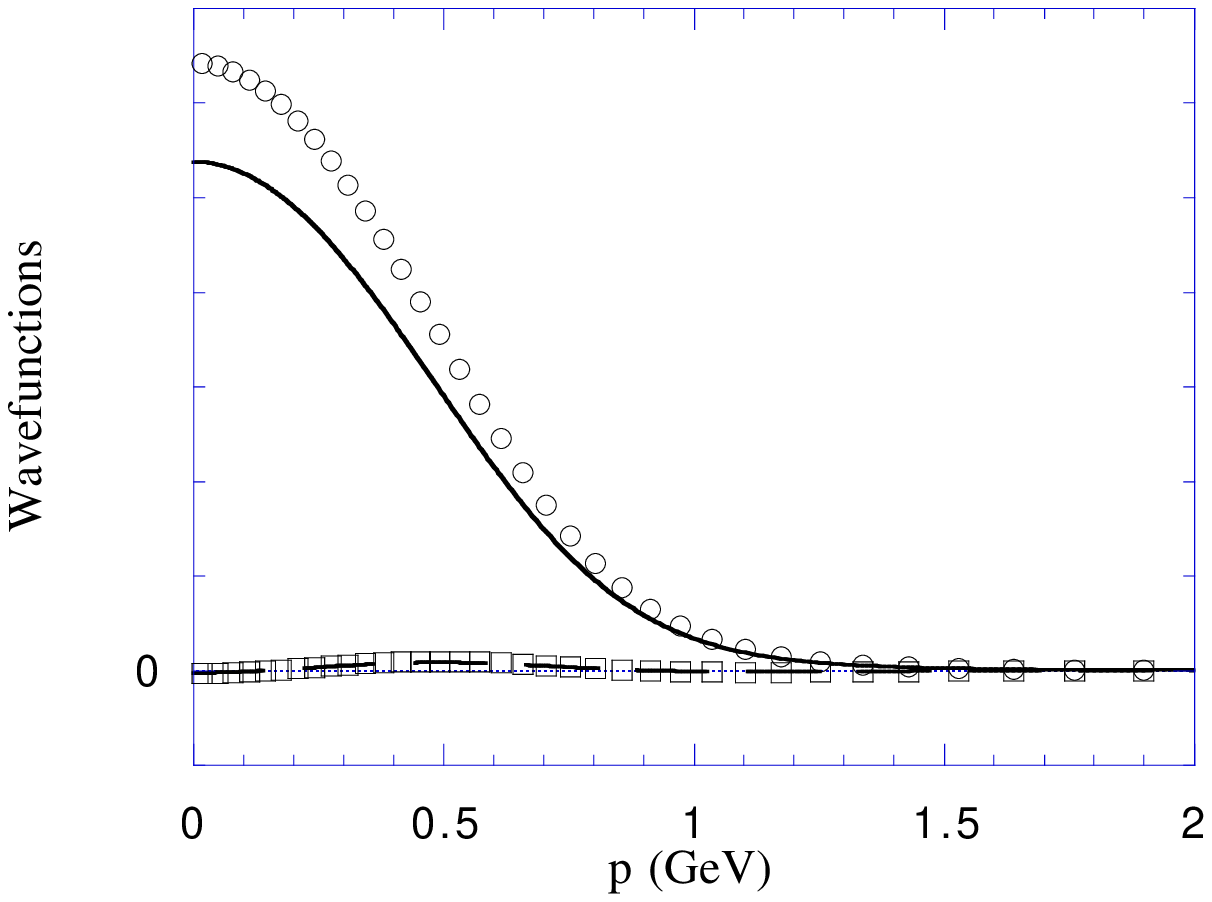}
}
\end{center}
\caption{Stable ground state solution for the Salpeter equation with
a {\it pure scalar\/} confining interaction.  In these cases $m=0.85$ GeV and
the Salpeter bound state has mass $\mu_{1}$=2.185 GeV.  The solutions for SN=20
(circles and squares) and SN=30  (solid and long dashed lines) are compared.  (Note
that the solutions are not normalized -- see the discussion in the text.) }
\label{fig29}
\end{figure}
%

While it is true that the {\it scalar\/} Salpeter equation is unstable for
equal quark-antiquark masses of 0.325 GeV, increasing the mass of
the quarks will give stable solutions (this was anticipated by the
discussion in Sec.~III).  We find that the lower mass states of the $y$=0.0 Salpeter
equation are stable when the quark mass is increased to $m$=0.85. The ground state
wave functions for this case are shown in Fig.~\ref{fig29}, where
solutions for spline ranks of 20 and 30 are compared (since the wave functions have
not been normalized, only the {\it shape\/} of the two solutions should be
compared).  Solutions obtained for somewhat lower masses (m=0.65, for example)
appear stable for SN=20, but the spectrum shows some instability for SN=30.  In
general, the number of stable states for the pure scalar Salpeter equation increases
as the quark mass increases.  Further study is needed to obtain a
detailed understanding of the stability of the purely scalar Salpeter equation.

\begin{table}
\begin{center}
\begin{minipage}{6.0in}
\caption{Stability Results (the table lists the region of stability or the
first of the four tests that the system fails).}\label{tableVI}
\begin{tabular}{lllll}
 & $y=0.0$ & $y=0.4$ & $y=0.6$ & $y=1.0$\\
\tableline\\
Dirac  & stable & stable & Cond 2 & Cond 2 \\[.1cm]
1CS $\kappa$=1.0 & positive & Cond 3 & Cond 2 & Cond 2 \\
1CS $\kappa$=2.0 & positive & Cond 3 & Cond 2 & Cond 2 \\
1CS $\kappa$=5.0 & stable & positive & Cond 2 & Cond 2 \\
1CS $\kappa$=10.0 & stable & stable & Cond 2 & Cond 2 \\[.1cm] 
Salpeter  & m$\ge$ 0.85 GeV & stable & stable & stable \\[.1cm]
\end{tabular}
\end{minipage}
\end{center}
\end{table}

\section{Conclusions}

Table~\ref{tableVI} summarizes the results presented in the previous
sections, which are also outlined below:

\begin{itemize}

\item The Dirac equation is stable if the scalar confinement is stronger
than the vector confinement ($y<1/2$).

\item The Salpeter equation is stable if the interaction is {\it  mostly vector\/},
and perhaps also for pure scalar exchanges with a large quark mass.   The precise
boundary between stable and unstable solutions is presumably a function of the quark
mass $m$ and the vector strength $y$, and we have not mapped it out. 

\item The one channel spectator (1CS) equation has the Dirac limit, as
expected.  This means that for large mass ratios $\kappa=m_1/m_2$, it is
stable if the interaction is predominately scalar ($y<1/2$).  However, as
the mass ratio decreases toward unity, the region of instability grows. 
As we decrease $\kappa$ for a fixed vector strength $y<1/2$, the negative
energy states will first become unstable, and then the positive energy
states may follow.  However, if the vector strength is small enough (e.g.,
$y=0$) the positive energy states appear to be stable for all mass ratios.

\end{itemize}

\noindent The usefulness of an equation where only part if the spectrum is
stable depends on whether or not the spectrum of unstable states is
clearly separate from the spectrum of stable states (i.e. Condition 3 is
met).  The 1CS equation for scalar confinement has this feature; the unstable
states are those which map, in the Dirac limit, into negative energy
states.   If one is content to exclude these states from consideration on physical
grounds then the {\it scalar\/} 1CS equation can be used to describe confined
$Q\bar{q}$ systems for all mass ratios.  The Salpeter equation can also be used for
equal mass $q\bar{q}$ systems unless the confinement is predominately scalar and the
quark masses are not large. 

This conclusion answers one of the questions raised in the introduction;
clearly the stability of vector or scalar confinement depends on the
relativistic equation used.  Scalar confinement is stable if the 1CS
equation is used and vector confinement is stable if the Salpeter equation
is used.

We emphasize that our study of the stability of the spectator equation is
preliminary for three reasons:

\begin{itemize}

\item Only the 1CS equation has been studied.  As emphasized before, 
a {\it two\/} channel spectator equation must be used if the bound state mass
is small (the pion), and any spectator equation must be explicitly
symmetrized if the quark masses are equal.  

\item Our study of the 1CS equation was limited to the quasirelativistic
approximation, in which retardation is neglected.  However, neglecting
retardation usually leads directly to the Salpeter equation, and the
attempt to include it (at least approximately) is the principle reason for
choosing to use a spectator equation in the first place.  Including
retardation in our analysis (planned for a later work)  may alter our
conclusions.

\item Only the time-like part of a vector confinement (i.e. $\gamma^0\gamma^0$)
has been studied.  There are preliminary indications that our results will change
when the full vector interaction $\gamma^\mu\gamma_\mu$ is included.  

\end{itemize}

Our results for the Salpeter equation agree with Ref.~\cite{munz}, but
disagree with the results obtained by Parramore and Piekarewicz \cite{pp},
who found the Salpeter equation to be unstable once the vector strength
dropped below one-half, regardless of quark mass.  However, as stated above,
we find that the Salpeter equation is stable for a vector strength 0.4, and
is even stable for a pure scalar interaction provided the quark mass is
sufficiently large.  We looked at one of the cases they found to be unstable 
($\sigma$=0.29, $m$=0.9 GeV, with 25 basis states), and found it to contain
stable states.  A possible explanation for this difference is that we use cubic
splines for our basis functions, while nonrelativistic harmonic oscillator wave
functions were used in Ref.~\cite{pp}.

There are other equations which can be used to model the quark-antiquark system.
Tiemeijer and Tjon \cite{TT} explored two such equations, the
Blankenbecler-Sugar-Logunov-Tavkhelidze (BSLT)\cite{BSLT} equation and the equal-time
(ET) equation of Wallace and Mandelzweig\cite{mw}.  The kernels for both
equations contained one-gluon-exchange (with the full
four vector structure) and a linear confining term (with a mixed
scalar-four vector structure). They found that increasing the
vector strength  of the confining term improved the phenomenology, but that some
mesons became unstable for vector strengths of more than about 0.25, depending on the
equation and gauge used.  These results reinforce the general conclusions of this
paper: stability depends on both the Lorentz structure of confinement and on the
type of relativistic equation used.

We have seen that the study of the mathematical stability of relativistic
equations requires the examination of both local and {\it global\/}
features of the eigenvalue spectrum and have introduced four conditions
which must be satisfied for an equation to give stable solutions.  Using
these stability criteria we find that the Lorentz
structure of the kernel and  the equation used to model the meson both play
a crucial role in the mathematical stability of the system.  Clearly further
research on this topic is needed.

\acknowledgments
The support of the DOE through grant  No.~DE-FG02-97ER41032 is gratefully
acknowledged.  We would also like to acknowledge a helpful conversation
with John Tjon, and one with Steve Wallace who called our attention to
new, related work on this subject\cite{wallace}.  In this work the
stability of the equal time equation of Wallace and Mandelzweig\cite{mw}
is examined, and they find results complementary to ours. 
\appendix
\section{Spline functions and numerical methods}

To solve the equations in this paper numerically, we expand each momentum
space wave function in terms of cubic splines
\begin{equation}
\psi_i(p)=\sum_{j=1}^{{\rm SN}}\alpha^i_{j}\beta_{j}(p) \, ,\label{iddr29}
\end{equation}
where $\alpha^i_{j}$ are the expansion coefficients (which become the
eignevectors of the problem), $\beta_{j}$ are the spline functions, and SN
is the number of spline functions in the expansion (the spline rank).  In all
of the equations studied there are only two independent wave functions, so
$i=$ 1 or 2. Since the angular integrations are performed analytically, the
wave functions depend only on the magnitude of the momentum  $p$.  Once
Eq.~(\ref{iddr29}) is substituted for each of the wave functions, both sides
of the  equation are operated on by the integral operator
\begin{equation}
\int p^{2}\beta_{l}(p)dp\, . \label{iddr30}
\end{equation}
This reduces the integral equations to matrix equations with dimension
2SN$\times$2SN and of the general form
\begin{eqnarray}
\left\{\lambda\pmatrix{A_{lj} & 0\cr 
                0 & A_{lj}} + \pmatrix{B^{11}_{lj} &  0\cr 
                0 & B^{22}_{lj}}
-\pmatrix{V^{11}_{lj} & V^{12}_{lj}\cr 
                V^{21}_{lj} & V^{22}_{lj}}\right\} \pmatrix {\alpha^1_j \cr
\alpha^2_j} =0 \, . \label{iddr31}
\end{eqnarray} 
These equations are then solved for the eigenvalues $\lambda$ and the
eigenvectors $\{\alpha^1_j,\;\alpha^2_j\}$.  In the following subsections
we give the forms of the matrices $A$ and $V$ for each case studied in
this paper.

\subsection{Dirac equation}

The Dirac equation was given in Eq.~(\ref{dirac1}) and (\ref{iddr28}).
The two independent wave functions are 
\begin{eqnarray}
\psi_1=&&\psi_{1a}\nonumber\\
\psi_2=&&\psi_{1b}\, , \label{wf1}
\end{eqnarray}
and $\lambda=E_B$, 
\begin{eqnarray}
A_{lj} = \int_0^\infty p^{2}dp\;\beta_{l}(p)\beta_{j}(p) \, , \label{A}
\end{eqnarray}
and 
\begin{eqnarray}
B^{11}_{lj} = - B^{22}_{lj} =- \int_0^\infty p^{2}dp\;
E_p\,\beta_{l}(p)\beta_{j}(p) \, .
\end{eqnarray}
Setting $m_2=m$ and using the notation
\begin{equation}
f_l(p)=\frac{N_{p}}{\sqrt{2E_{p}}}\;\beta_{l}(p)\, ,
\end{equation}
the potential matrix can be written 
\begin{eqnarray}
&&\pmatrix{V^{11}_{lj} & V^{12}_{lj}\cr 
                V^{21}_{lj} & V^{22}_{lj}} = \nonumber\\
&&\qquad-\frac{4\sigma}{\pi}\int_0^\infty\int_0^\infty
dp\,dk\;V_0(p,k)\,f_{l}(p)
\left\{f_{j}(k)\pmatrix{\eta_1 & \eta_2 \cr \eta_3 &
\eta_4}-f_{j}(p) \pmatrix{\eta'_1 & \eta'_2 \cr
\eta'_3 & \eta'_4}\right\}
 \nonumber\\
&&\qquad-\frac{4\sigma}{\pi}\int_0^\infty\int_0^\infty
dp\,dk\;V_1(p,k)\,f_{l}(p)
f_{j}(k)\pmatrix{\zeta_5 & \zeta_6 \cr \zeta_7 & \zeta_8}
\, . \label{iddr31a}
\end{eqnarray}
The functions $\eta$ and $\zeta$ are
\begin{equation}
\eta_i=a_i+b_i \qquad \zeta_i=b_i
\end{equation}
where $a_i$ and $b_i$ were defined in Eq.~(\ref{iddr26a}), and if
$\eta_i=\eta_i(p,k)$, then
$\eta'_i=\eta_i(p,p)$.  The functions $V_0$ and $V_1$ are
\begin{eqnarray}
V_0(p,k)=\frac{1}{2}\int_{-1}^1dz\frac{p^2k^2}{(p^2+k^2-2pkz)^2}=&&
\frac{p^2k^2}{(p^2+k^2)^2-4p^2k^2} \nonumber\\ 
V_1(p,k)=\frac{1}{2}\int_{-1}^1dz\frac{p^2k^2\,(z-1)}{(p^2+k^2-2pkz)^2}=&&
\frac{1}{2}
\frac{pk}{(p^2+k^2+2pk)}\nonumber\\
&& - \frac{1}{8} \log \left(
\frac{p^2+k^2+2pk}{p^2+k^2-2pk}\right) \, .
\end{eqnarray}

\subsection{One-channel spectator equation}

The 1CS equation in helicity form was given in Eq.~(\ref{der23}) with 
the potential defined in  Eq.~(\ref{iddr28rel}) [with $(p-k)^2\to({\bf
p}-{\bf k})^2$ as discussed in Sec.~IIB].  The two independent wave
functions are as in Eq.~(\ref{wf1}) and $\lambda=\mu=m_1+E_B$.  The matrix
$A$ is identical to the Dirac case, but now
\begin{eqnarray}
B^{11}_{lj} &&= -\int_0^\infty p^{2}dp\;
\left(E_{p_1}+E_{p_2}\right)\,\beta_{l}(p)\beta_{j}(p) \nonumber\\
B^{22}_{lj} && = -\int_0^\infty p^{2}dp\;
\left(E_{p_1}-E_{p_2}\right)\,\beta_{l}(p)\beta_{j}(p)\, .
\end{eqnarray}
Introducing the notation
\begin{equation}
F_l(p)=\frac{N_{p_1}N_{p_2}}{\sqrt{4E_{p_1}E_{p_2}}}\;\beta_{l}(p)\, ,
\end{equation}
the potential matrix can be written
\begin{eqnarray}
&&\pmatrix{V^{11}_{lj} & V^{12}_{lj}\cr 
                V^{21}_{lj} & V^{22}_{lj}}=\nonumber\\
&&\qquad-\frac{4\sigma}{\pi}\int_0^\infty\int_0^\infty
dp\,dk\;V_0(p,k)\,F_{l}(p)
\left\{F_{j}(k)\pmatrix{\overline{\eta}_1 & \overline{\eta}_2 \cr
\overline{\eta}_3 &
\overline{\eta}_4}-\frac{E_{p_1}}{E_{k_1}}F_{j}(p)
\pmatrix{\overline{\eta}'_1 &
\overline{\eta}'_2 \cr \overline{\eta}'_3 & \overline{\eta}'_4}\right\}
 \nonumber\\
&&\qquad-\frac{4\sigma}{\pi}\int_0^\infty\int_0^\infty
dp\,dk\;V_1(p,k)\,F_{l}(p)
F_{j}(k)\pmatrix{\overline{\zeta}_1 & \overline{\zeta}_2 \cr 
\overline{\zeta}_3 & \overline{\zeta}_4}
\, , \label{a5}
\end{eqnarray}
with 
\begin{equation}
\overline{\eta}_i=A_i+B_i \qquad \overline{\zeta}_i=B_i\, ,
\end{equation}
where the $A_i$ and $B_i$ were defined in Eq.~(\ref{der25b}).
The meaning of the prime in $\overline{\eta}'$ is the same as in $\eta'$ and
$V_0$ and $V_1$ are as before.

\subsection{Salpeter equation}

The Salpeter equation is given in Eq.~(\ref{spder2a}), with the masses
both equal to $m$.  Now
\begin{eqnarray}
\psi_1=&&\psi_{1a}\nonumber\\
\psi_2=&&\psi_{2a}\, , \label{wf2}
\end{eqnarray}
and $\lambda=\mu$.  The matrix $A$ is identical, but $B$ is two times 
the Dirac $B$.  The potential matrix is similar to Eq.~(\ref{a5}) with 
\begin{equation}
\begin{array}{ll}
\overline{\eta}_2\to \eta_5 \qquad &\overline{\zeta}_2\to
\zeta_5\cr
\overline{\eta}_3\to -\eta_5 \qquad &\overline{\zeta}_3\to -\zeta_5\cr
\overline{\eta}_4\to -\overline{\eta}_1 \qquad &\overline{\zeta}_4\to
-\overline{\zeta}_1\, ,
\end{array}
\end{equation}
and, from Eq.~(\ref{spder3}),
\begin{eqnarray}
\eta_5=\tilde{p}^2+\tilde{k}^2 + 2(1-2y)\tilde{p}\tilde{k}\qquad
\zeta_5=2(1-2y)\tilde{p}\tilde{k}\, .
\end{eqnarray}

\subsection{Splines}

The solution to the wavefunctions used in this paper are based on a set of 
third order polynomial functions called cubic splines.  Used previously in
papers such as  Ref.~\cite{gm2}, they have proven versatile enough to model
all of the wavefunctions examined in this paper.  

\begin{figure}
\begin{center}
\mbox{
   \epsfysize=3.0in
\epsfbox{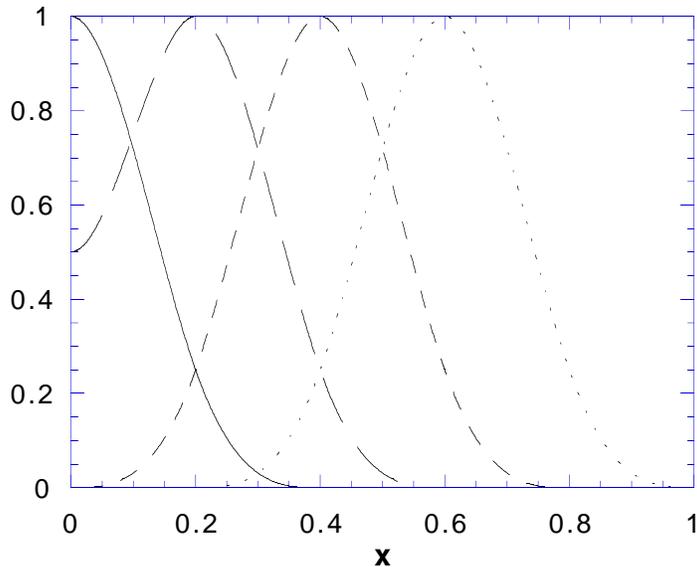}
}
\end{center}
\caption{Spline rank 4 curves (1 solid, 2 long dashed, 3 short dashed, and 4
dotted).}
\label{spline1}
\end{figure}

The wave function expansion was given in Eq.~(\ref{iddr29}).  Each spline is 
constucted from four separate functions.  The function used depends on the
argument and the spline index $j$ as shown:    
\begin{equation}
4\beta_{j}(x)=\left\{\begin{array}{ll} \frac{(x-x_{j-2})^3}{h^3}\, , & x\in
[x_{j-2},x_{j-1}]\\ 
1+3\frac{(x-x_{j-1})}{h}+3\frac{(x-x_{j-1})^2}{h^2}-3\frac{(x-x_{j-1})^3}
{h^3}\, , & x\in [x_{j-1},x_{j}] \\ 
1+3\frac{(x_{j+1}-x)}{h}+3\frac{(x_{j+1}-x)^2}{h^2}-3\frac{(x_{j+1}-x)^3}{h^3}\, , &
x\in [x_{j},x_{j+1}] \\ 
\frac{(x_{j+2}-x)^3}{h^3}\, , & x\in [x_{j+1},x_{j+2}]\\ 
0 & {\rm otherwise}\, . \\ 
\end{array}\right.
\label{appd2}
\end{equation}  
Each spline is defined on the interval from zero to one.  This range
is divided into sectors whose size, $h=1/({\rm SN}+1)$, depends on the spline
rank.  Each sector is bounded by nodes at $x_k$ and $x_{k+1}$, with the
number of nodes equal to SN$+2$.  The first node, $x_1$, is always located
at zero, and the last one, $x_{{\rm SN}+2}$, at one.  The spline curves
for a spline rank of 4 are given in Fig.~\ref{spline1}.  The standard
choice for our calculation was a rank of 20 (20 splines in
each wave function expansion).

None of the nodes may lie outside of the interval from 0 to 1, so the first
spline, $j=1$, is defined entirely by the third and forth functions given in
Eq.\,(\ref{appd2}).  It has a zero slope at $x=0$.  The $j=2$ spline was
defined in a special way so that it too will have zero slope at
$x=0$ (insuring that all the splines have this property).  To accomplish 
this the first sector (which lies between
$x_{0}$ and $x_1$ and is hence outside the acceptable range of support) will
be ``folded over'' onto the interval between $[x_1,x_2]$.  Hence, {\it in the
interval between\/} $[x_1,x_2]$ the second spline is defined to be
\begin{equation}
4\beta_2(x)=1+3\frac{(x-x_{1})}{h}+3\frac{(x-x_{1})^2}{h^2}-
3\frac{(x-x_{1})^3} {h^3} + \frac{(x_{2}-x)^3} {h^3} \, .
\end{equation}
This is an exceptional case, and all other splines are defined following
Eq.~(\ref{appd2}) in a straightforward fashion.

\begin{figure}
\begin{center}
\mbox{
   \epsfysize=3.0in
\epsfbox{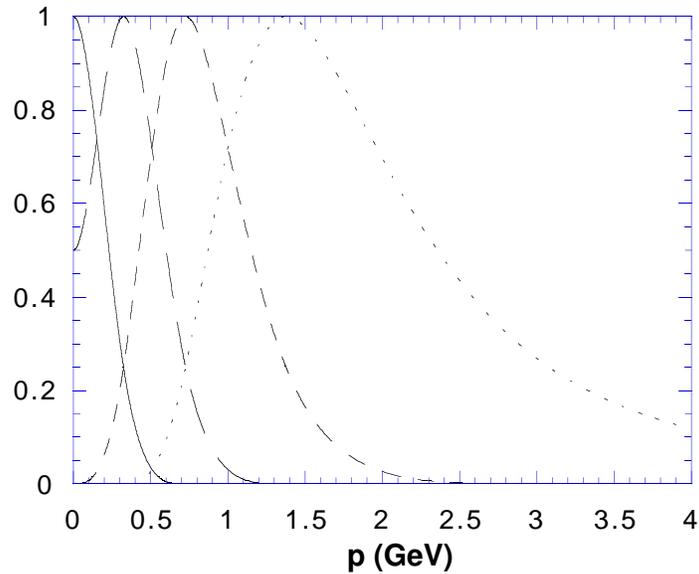}
}
\end{center}
\caption{Spline rank 4 curves (1 solid, 2 long dashed, 3 short dashed, and 4
dotted) with momentum arguement.}
\label{spline2}
\end{figure}

The splines defined in Eq.~(\ref{appd2}) are only continuous up to their
second derivative.  Therefore, in order to obtain convergence the integrals
must be separately evaluated for each sector,  and the results from
all the sectors summed up afterwards.  Special care must be taken in
evaluating those contributions to the double integral of the potential
which include singularities.  These are evaluated by choosing points
equally spaced on each side of the singularity so that a well defined limit
is obtained.

To use the splines to describe the wave functions, the interval
$[0,\infty)$ is mapped into the line segment $[0,1]$ using the tangent
mapping
\begin{equation}
x=\frac{2}{\pi}\arctan\left(\frac{p}{\Lambda}\right)\, ,
\label{appd3}
\end{equation}    
with $\Lambda=1$ GeV. This mapping alters the shape of the splines, as
illustrated in Fig.~\ref{spline2}.  

When the spline rank is
increased the sectors become smaller and the range in momentum space over
which the splines are significantly different from zero increases.  Thus,
the wave function is more accurately modeled as the spline rank increases. 
Of course this higher precision must be balanced by consideration of
computation time.    

\end{document}